\documentclass[twocolumn,10pt,aps,pra,showpacs,superscriptaddress,nobalancelastpage,longbibliography,nofootinbib,floatfix]{revtex4-2}
\usepackage{graphicx}
\usepackage{dcolumn}
\usepackage{bm}
\usepackage{braket} 
\usepackage{amssymb}
\usepackage{amsmath}

\usepackage{algorithm}
\usepackage[noend]{algpseudocode}
\usepackage{graphicx} 
\usepackage{subcaption}
\usepackage{wrapfig}
\usepackage{adjustbox}
\usepackage{relsize}
\usepackage{hyperref}
\usepackage{xcolor}
\usepackage{svg}
\usepackage{natbib}
\usepackage{multirow}
\usepackage{afterpage}
\usepackage{tikz} 
\usetikzlibrary{quantikz} 
\usepackage{hhline}

\begin{document}

\title{Transversal CNOT gate with multi-cycle error correction}

\author{Younghun Kim}
\email{younghunk@student.unimelb.edu.au}
\affiliation{School of Physics, The University of Melbourne, Parkville, 3010, Victoria, Australia}
\affiliation{Data61, CSIRO, Clayton, 3168, Victoria, Australia}

\author{Martin Sevior}
\email{martines@unimelb.edu.au}
\affiliation{School of Physics, The University of Melbourne, Parkville, 3010, Victoria, Australia}

\author{Muhammad Usman}
\email{muhammad.usman@unimelb.edu.au}
\affiliation{School of Physics, The University of Melbourne, Parkville, 3010, Victoria, Australia}
\affiliation{Data61, CSIRO, Clayton, 3168, Victoria, Australia}

\begin{abstract}
A scalable and programmable quantum computer holds the potential to solve computationally intensive tasks that classical computers cannot accomplish within a reasonable time frame, achieving quantum advantage. However, the vulnerability of the current generation of quantum processors to errors poses a significant challenge towards executing complex and deep quantum circuits required for practical problems. Quantum error correction codes such as Stabilizer codes offer a promising path forward for fault-tolerant quantum computing, however their realisation on quantum hardware is an on-going area of research. In particular, fault-tolerant quantum processing must employ logical gates on logical qubits with error suppression with realistically large size codes. This work has implemented a transversal CNOT gate between two logical qubits constructed using the Repetition code with flag qubits, and demonstrated error suppression with increasing code size under multiple rounds of error detection. By performing experiments on IBM quantum devices through cloud access, our results show that despite the potential for error propagation among logical qubits during the transversal CNOT gate operation, increasing the number of physical qubits from 21 to 39 and 57 can suppress errors, which persists over 10 rounds of error detection. Our work establishes the feasibility of employing logical CNOT gates alongside error detection on a superconductor-based processor using current generation quantum hardware.

\end{abstract}

\maketitle
\section{Introduction}
Noisy Intermediate-Scale Quantum (NISQ) \cite{NISQ} devices consist of several dozen to a few hundred noisy physical qubits accessible through various quantum processor platforms, such as superconductors \cite{heavy-hexagon,sycamore}, trapped ions \cite{quantinuum,quantinuum2} and neutral atoms \cite{neutral}. Despite the potential of quantum processors to transcend classical computers in computational capabilities \cite{sycamore}, solving practical problems by interpreting results from complex quantum circuits remains challenging due to errors. To obtain reliable results, it is crucial to limit errors below a target error rate \cite{rsa}, which can be achieved by constructing a fault-tolerant quantum computer with logical qubits and gates protected against errors by quantum error correction codes, such as those provided by Stabilizer codes \cite{gottesman,surface,qecmemory}. Stabilizer codes allow the implementation of arbitrarily accurate quantum computation, provided that the error rate per physical qubit gate is kept below a certain acceptable threshold value \cite{threshold,threshold2,threshold3}.

\begin{table*}[htbp]
\caption{ Lists of various error correction experiments with logical gates. Quantum error correction codes are denoted as $[[n,k,d]]$, where $n$, $k$, and $d$ represent the number of data qubits, logical qubits, and the distance. Codes that can protect against either X or Z errors are denoted as $[n,k,d]$. The table indicates the number of syndrome extraction rounds, the hardware type used in each experiment, and error suppression achievement by increasing the size of the code block. Additionally, the table specifies the type of logical transversal gates utilized. }
\resizebox{\textwidth}{!}{
\begin{tabular}{ |c | c | c | c | c | c | c | c| }
\hline
\multirow{2}{*}{Reference} & \multirow{2}{*}{Year} & \multirow{2}{*}{Structure} & \multirow{2}{*}{Logical Gates} & Error Suppression & \multirow{2}{*}{Rounds} & Hardware \\ 
 & & & & Demonstration & & Modality \\ \hline

\multirow{2}{*}{\cite{trapped_ion_transversal}} & \multirow{2}{*}{2022}
 & \multirow{2}{*}{$[[7,1,3]]$} & \multirow{2}{*}{CNOT} & \multirow{2}{*}{No} & \multirow{2}{*}{Single} & \multirow{2}{*}{Trapped-ion} \\
 & & & & & & \\ \hline

\multirow{2}{*}{\cite{entangling_gate}} & \multirow{2}{*}{2022}
 & $[[10,2,3]]$, & \multirow{2}{*}{CNOT} & \multirow{2}{*}{No} & \multirow{2}{*}{Single} & Trapped-ion \\
 & & $[[7,1,3]]$ & & & & (Quantinuum) \\ \hline
 
\multirow{2}{*}{\cite{ibm_transversal}} & \multirow{2}{*}{2023} & \multirow{2}{*}{$[[8,3,2]]$}
 & CZ & \multirow{2}{*}{No} & \multirow{2}{*}{Single}  & Trapped-ion (IonQ) \\
 & & & CCZ & & & Superconductors (IBM) \\ \hline
 
\multirow{2}{*}{\cite{ion_cnot}} & \multirow{2}{*}{2023}
 & $[[9,1,3]]$-$[[49,1,7]]$, & CNOT, CZ, & \multirow{2}{*}{Yes} & \multirow{2}{*}{Single} & Neutral-atom\\
 & & $[[8,3,2]]$ & CCZ & & & (QuEra)\\ \hline
 
\multirow{2}{*}{\cite{carbon}} & \multirow{2}{*}{2024}
 & $[[7,1,3]]$, & \multirow{2}{*}{CNOT} & \multirow{2}{*}{Yes} & \multirow{2}{*}{Single} & Trapped-ion \\
 & & $[[12,2,4]]$ & & & & (Quantinuum) \\ \hline

\multirow{2}{*}{\cite{heavy-hex-entangle}} & \multirow{2}{*}{2024} & \multirow{2}{*}{$[[4,1,2]]$-$[[16,1,4]]$}
 & \multirow{2}{*}{CNOT} & \multirow{2}{*}{No} & \multirow{2}{*}{Five} & Superconductors \\
 & & & & & & (IBM) \\ \hline

\multirow{2}{*}{This work} & \multirow{2}{*}{2024} & \multirow{2}{*}{$[3,1,3]$-$[7,1,7]$}
 & \multirow{2}{*}{CNOT} & \multirow{2}{*}{Yes} & \multirow{2}{*}{Ten} & Superconductors \\
 & & & & & & (IBM) \\ \hline
 
\end{tabular}}
\label{table1}
\end{table*}

Significant advances in the development of quantum devices have enabled the field of quantum error correction to shift from purely theoretical analysis to hardware benchmarking, with several recent experiments reporting the principles and realization of Stabilizer codes employing various structures, such as the Repetition code \cite{MWPM,ibm_repetition,google_qec1,repetition_flag}, Surface code \cite{google_qec2,d2_qec,d3_surf,d3_surper,hardware_err}, and Color code \cite{d3_ion,d3_ion2}. Experimental results have shown that by increasing the number of physical qubits, a logical qubit can be constructed on quantum devices offering tolerance against errors with feedback based on detected errors over several rounds of the error detection circuit, also known as a syndrome extraction circuit \cite{google_qec1,repetition_flag,google_qec2}. However, for practical quantum computing, it is essential to demonstrate the implementation and scaling of logical gates on logical qubits \cite{outlook} -- a feat which has not been fully achieved yet.


Amongst multi-qubit logical operations, the implementation of a two-qubit logical CNOT gate on quantum hardware will be an essential step towards achieving universal quantum gate set \cite{universality,universality2, universal,transversal_restrict} and fault-tolerant quantum computing. One approach to realizing the logical CNOT gate is by performing a transversal CNOT gate \cite{transversality,transversality2,transversality3,transversality4}. Table \ref{table1} summarizes previous studies investigating the performance of multi-logical qubit gates including the transversal CNOT gate on real quantum devices \cite{trapped_ion_transversal,entangling_gate, ibm_transversal, ion_cnot, carbon, heavy-hex-entangle}, noting that all of these studies have been reported within the last couple of years indicating that the research on multi-qubit logical gate implementation is still emerging. These studies have implemented multi-logical qubit entangled states, like the logical Bell state, with a primary focus on evaluating their performance in terms of state preparation. Furthermore, previous experimental literature has explored quantum teleportation at the logical qubit level using transversal CNOT gates \cite{teleportation}. However, none of these papers have demonstrated error suppression of the logical CNOT gate with multiple cycles of syndrome extraction circuits through scaling up the code block size. Therefore, investigating the performance of a logical CNOT gate and demonstrating the possibility of error suppression while tracking errors on real quantum computers remains an open question in the NISQ era.

\begin{figure}[h]
    \centering
    \includegraphics[width=0.5\textwidth]{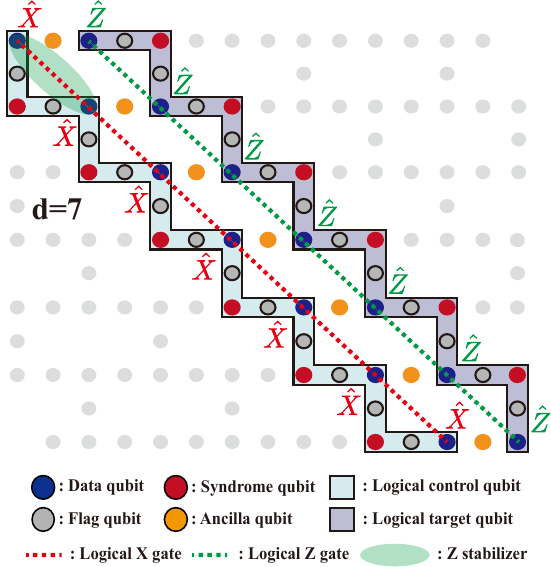}
    \caption{ Mapping the structures for $d=7$ onto IBM's hardware. The transversal gate is designed to implement the logical CNOT gate on two logical qubits: the logical control (sky blue area) and the target qubit (purple area). Each logical qubit has data, flag, and syndrome qubits corresponding to the blue, grey, and red dots, respectively. The structure can be mapped onto IBM's quantum hardware (127 physical qubits) for different distances $d\in \{3,5,7\}$, corresponding to structures with 21, 39, and 57 physical qubits including ancilla qubits for implementing the transversal CNOT gate. }
    \label{fig1}
\end{figure}

In this work, we design a quantum memory experiment to assess the effectiveness of a transversal logical CNOT gate, including multiple rounds of a syndrome extraction circuit. To evaluate the performance of a transversal CNOT gate, we prepare two logical qubits using the Repetition code with flag qubits \cite{repetition_flag}. The quantum memory experiment consists of multiple rounds of syndrome extraction circuits, with the transversal CNOT gate placed between these rounds. We utilize ancilla qubits to implement the transversal CNOT gate and map the structure onto a heavy-hexagon configuration. Our experiments are conducted on the \texttt{ibm\_sherbrooke} and \texttt{ibm\_torino} quantum devices \cite{torino}, which support structures ranging from 21 to 57 physical qubits, as illustrated in Figure \ref{fig1}. We collect sampled data from the devices by executing the designed quantum circuit and then use a decoder to correct the final states of the two logical qubits. Logical error rates are computed by the expectation values of the logical Pauli gates known as the infidelity. Our results demonstrate that although error propagation among logical qubits can occur with the transversal CNOT gate, increasing the number of physical qubits can suppress errors by bit- or phase-flip error detection and correction on one of the IBM quantum processors. The error suppression remains consistent across ten cycles of the syndrome extraction circuit. 

The paper begins by explaining the Repetition code, which utilizes flag qubits, along with its syndrome extraction circuit. Next, we illustrate the transversal CNOT gate employing ancilla qubits between two logical qubits, detailing its structure and discussing the quantum memory experiment for sampling data on the devices. Subsequently, we present a syndrome graph for error correction in the sampled data. Additionally, we visualize errors using a correlation matrix based on the data. As a result, we demonstrate the logical error probability of the transversal CNOT gate as a function of the number of physical qubits and the rounds of the syndrome extraction circuit. Finally, we summarize our findings and conclude.

\section{Method}
\subsection{Repetition with flag qubits}\label{repetition}

\begin{figure*}[htbp]
    \centering
    \includegraphics[width=\textwidth]{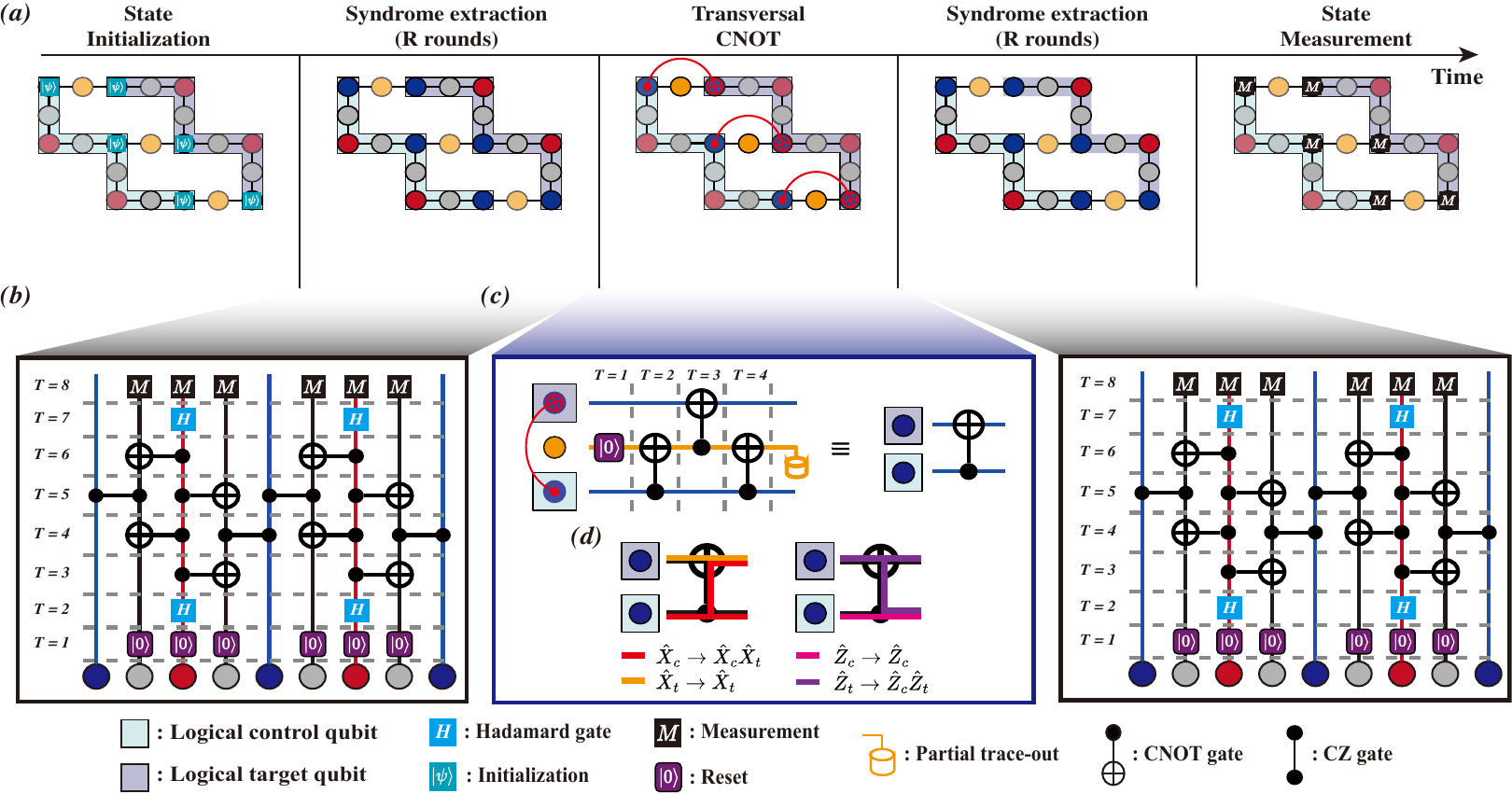}
    \caption{ Quantum memory experiment with the transversal CNOT gate. (a) A quantum circuit used for sampling data in the case of 21 physical qubits structure is depicted. The quantum circuit has five steps: First, two logical qubits are encoded via data qubit initialization. (b) Secondly, the syndrome extraction circuit is implemented \textbf{R} times for each logical qubit with eight steps. The subroutine circuit creates interaction between data qubits and, flag and syndrome qubits to detect errors from their measurement outcomes. (c) Next, the transversal CNOT gate is executed with three physical CNOT gate layers employing ancilla qubits. The CNOT gates are applied only between data qubits from two logical qubits and the ancilla qubits between them. (d) While physical qubits undergo a physical CNOT gate, an X or Z Pauli gate can propagate from one physical qubit to another. Subsequently, another \textbf{R} round of the syndrome extraction circuit is considered. Finally, every data qubits are measured in the Z basis. }
    \label{fig2}
\end{figure*}

Physical qubits used in Stabilizer codes can be categorized as data and syndrome qubits. While data qubits are used to encode quantum information, syndrome qubits detect errors. Stabilizer operators can be defined based on the Pauli operators of data qubits, and a syndrome extraction circuit is built using these stabilizers \cite{syndrom_extraction,syndrom_extraction_2,syndrom_extraction_3,qec_guide}. Additionally, flag qubits can be employed in the syndrome extraction circuit \cite{flag1,flag2}. An additional flag qubit can be introduced to detect multi-qubit errors in the data resulting from a single fault during the execution of the syndrome extraction circuit \cite{trapped_ion_transversal,entangling_gate,teleportation,d3_ion}, or to address hardware limitations in connectivity \cite{ibm_qec1,ibm_qec2,hh-code,strategy,heavyhexagon,heavyhexagon}, such as IBM's heavy-hexagon quantum hardware structure \cite{heavy-hexagon}. 

We construct a logical qubit using the Repetition code with flag qubits. A single logical qubit consists of data, syndrome, and flag qubits, depicted in Figure \ref{fig1}. All dots represent physical qubits; these include blue, red, and grey dots corresponding to data, syndrome, and flag qubits. The data and syndrome qubits are alternately placed but physically segregated, with a flag qubit filling the gap between them. The data qubits are encoded in the quantum state of a single logical qubit. Their state carries the observable information of interest and is subject to change through a logical gate. On the other hand, flag and syndrome qubits detect errors that may affect the state of data qubits. When the number of data qubits is $d$, known as a distance for the Repetition code, the structure requires $d-1$ syndrome and $2(d-1)$ flag qubits. 

Protecting a logical qubit against errors requires a stabilizer group constructed by a series of the Pauli operators of data qubits. The stabilizer group constrains the state of the data qubits to a logical state ($\ket{\psi}_L$) and enables the detection of errors that could lead the state to deviate from the logical state. The logical Pauli operators ($\hat{X}_L, \hat{Z}_L$) commute with all stabilizers. Hence, the logical Pauli operators change the state of the logical qubit without causing an error. A simple way to define them is by applying Z(X) operators on all data qubits, thereby implementing the logical Z(X) gate. 

In Figure \ref{fig1}, the stabilizers can be defined by the Z Pauli operators of a pair of neighbor data qubits as $\hat{Z}_i\hat{Z}_{i+1}$, where $i$ denotes a data qubit number. When a logical qubit contains three data qubits, the logical qubit can be protected from a bit-flip error and defined in the logical states ($\ket{0}_L=\ket{000}, \ket{1}_L=\ket{111}$). The logical Pauli operators are chosen as $\hat{X}_L=\hat{X}_1 \hat{X}_2 \hat{X}_3$ and $\hat{Z}_L=\hat{Z}_1 \hat{Z}_2 \hat{Z}_3$. In our work, we use X stabilizers to prepare the logical state in the X basis, which can detect and correct phase-flip errors that cause changes in its logical state. More details of the X stabilizers can be found in Appendix \ref{app:supp3.1}. 

A quantum circuit subroutine known as a syndrome extraction circuit detects errors in a logical qubit. The syndrome extraction circuit is designed and performed based on the stabilizers. The syndrome extraction circuit using Z stabilizers is constructed as Figure \ref{fig2} (b): First, syndrome qubits are prepared as $\ket{+}$ through the application of Hadamard gates, and flag qubits are initialized as $\ket{0}$ ($T=1, 2$). Secondly, each physical qubit is involved in a two-qubit gate configuration with its neighbor qubit, and the gates are applied according to a particular sequence in parallel. Two-qubit gates are only implemented between a single physical qubit and its nearest adjacent qubit at a hardware level ($T=3\sim6$). Finally, we measure every flag and syndrome qubits in the Z basis ($T=7, 8$). The syndrome extraction circuit can be repeated multiple times and map their outcomes to a binary vector, a syndrome, used as input of a decoder algorithm.

In the absence of errors, the measurement outcomes from flag and syndrome qubits after syndrome extraction circuits are all $\ket{0}$ states, consistently yielding the same values across multiple rounds. In this case, the syndrome is a zero vector. However, each component operation, including measurements, the initialization to the $\ket{0}$ state, as well as the one- (H) and two- (CNOT and CZ) qubit gates, are susceptible to errors during execution on quantum devices. Depending on the errors that occur as a quantum device executes syndrome extraction circuits, they flip the measured outcomes of either flag or syndrome qubits, changing them from $\ket{0}$ to $\ket{1}$ or vice versa. If measured outcomes of qubits are flipped compared to the previous round and have odd parity, it is considered a detection event resulting in a ``1" bit in the syndrome \cite{MWPM}. Depending on the errors that occur during the execution of quantum gates, the syndrome maps onto a corresponding binary vector.

\subsection{Quantum memory experiment}

The transversal CNOT gate is applied to two logical qubits: a logical control and a target qubit. The two logical qubits are prepared with the Repetition codes employing flag qubits. In Figure \ref{fig2} (a), the sky blue and purple boxes represent the logical control and target qubit. Each logical qubit has three data qubits, and the same number of ancilla qubits, denoted by the orange dots. 

The quantum memory experiment with the transversal CNOT gate is as follows: First, we encode the two logical qubits by initializing data qubits in either the Z or X basis depending on the given logical states. Second, we implement \textbf{R} rounds of the syndrome extraction circuit corresponding to the basis of the logical qubit, following the quantum gate sequences shown in Figure \ref{fig2}(b) for each logical qubit, as discussed in Method \ref{repetition}. Next, we use ancilla qubits and apply the transversal CNOT gate. Following the execution of the transversal CNOT gate, we perform the syndrome extraction circuit for \textbf{R} rounds. In total, the syndrome extraction circuit is executed \textbf{2R} rounds. Finally, we measure all data qubits in the Z basis. The structure for the transversal CNOT gate can be mapped onto the heavy hexagon structure and implemented on an IBM device for experimentation. When each logical qubit consists of $d$ data qubits, the structure requires $3d+6(d-1)$ physical qubits, including data, syndrome, flag, and ancilla qubits. 

The method for implementing the transversal CNOT gate is to execute a physical CNOT gate between locally corresponding data qubits. The physical CNOT gate is executed with the data qubit from the logical control qubit serving as the control qubit and the data qubit from the other logical qubit serving as the target qubit. The transversal CNOT gate implements the physical CNOT gate within each pair of grouped data qubits directly. Figure \ref{fig2} (a) shows two data qubits, the closest from each logical qubit, and the execution of a series of the physical CNOT gates for the transversal CNOT gate in parallel. However, owing to the heavy hexagon structure of IBM hardware, the data qubits from each logical qubit are not the nearest neighbor but have a physical qubit between them. To circumvent this issue, a physical qubit is used as an ancilla qubit to enable paired data qubits to interact for the transversal CNOT gate. 

To implement a transversal CNOT gate, three physical CNOT gates are applied. Figure \ref{fig2} (c) shows the sequence of quantum gates for implementing the transversal CNOT gate, specifying the order of physical CNOT gates. First, we initialize an ancilla qubit as $\ket{0}$ ($T=1$). Next, we apply the physical CNOT gate by choosing a data qubit of the logical control qubit as a control qubit and the ancilla qubit as a target qubit ($T=2$). Another physical CNOT gate is applied between the ancilla qubit and a data qubit of the logical target qubit ($T=3$). The data qubit of the logical control qubit and the ancilla qubit are selected for the control and target qubit, and we execute the last physical CNOT gate ($T=4$). The ancilla qubit is partially traced out after the physical CNOT gates have been implemented. The state of two data qubits after the series of quantum gates is the same as that resulting from the direct execution of a physical CNOT gate between them ideally \cite{heavy-hex-entangle,strategy}. We implement the physical CNOT gates individually and in parallel.

When a physical CNOT gate is applied, a physical Pauli operator can be transferred as shown in Figure \ref{fig2} (d). For an X gate on the control qubit before the physical CNOT gate, the resulting circuit is equivalent to X gates applied to both the control and target qubits after the CNOT gate (red line). Conversely, if the gate occurs on the target qubit, it remains isolated and does not affect the other qubit (orange line). In the case of a Z gate on the control qubit, there is no gate propagation (pink line), whereas the gate on the target qubit can lead to two Z gates applied to both qubits after the CNOT gate (purple line).

The transversal CNOT gate is designed with a series of physical CNOT gates and enables the propagation of X or Z gates between each paired data qubit. Hence, the logical Pauli operators ($\hat{X}_L$, $\hat{Z}_L$) formed using the Pauli operators of data qubits transversally follow the same gate propagation rules as their physical-level counterparts during the execution of the transversal CNOT operation at the logical level. However, a bit- or phase-flip error on a data qubit before the transversal CNOT gate can introduce an additional error on the interacted data qubit. In other words, the transversal CNOT gate also allows for the propagation of errors among the logical qubits. 

Correlated errors can arise and lead to ambiguity in calculating a correction operator by a decoder. This occurs not only because of errors within quantum hardware as it executes a quantum circuit \cite{correlated_err,hardware_err,leakage_err}, but also due to error propagation between two logical qubits through the transversal CNOT gate \cite{transversal_restrict,decoding_cnot,ion_cnot,heavy-hex-entangle}. Throughout this paper, we use the term ``Gate-Flow errors" to refer to errors caused by the transversal CNOT gate, distinguishing them from correlated errors inherent to the quantum hardware.

\subsection{Syndrome graph}

\begin{figure*}[t]
    \centering
    \includegraphics[width=\textwidth]{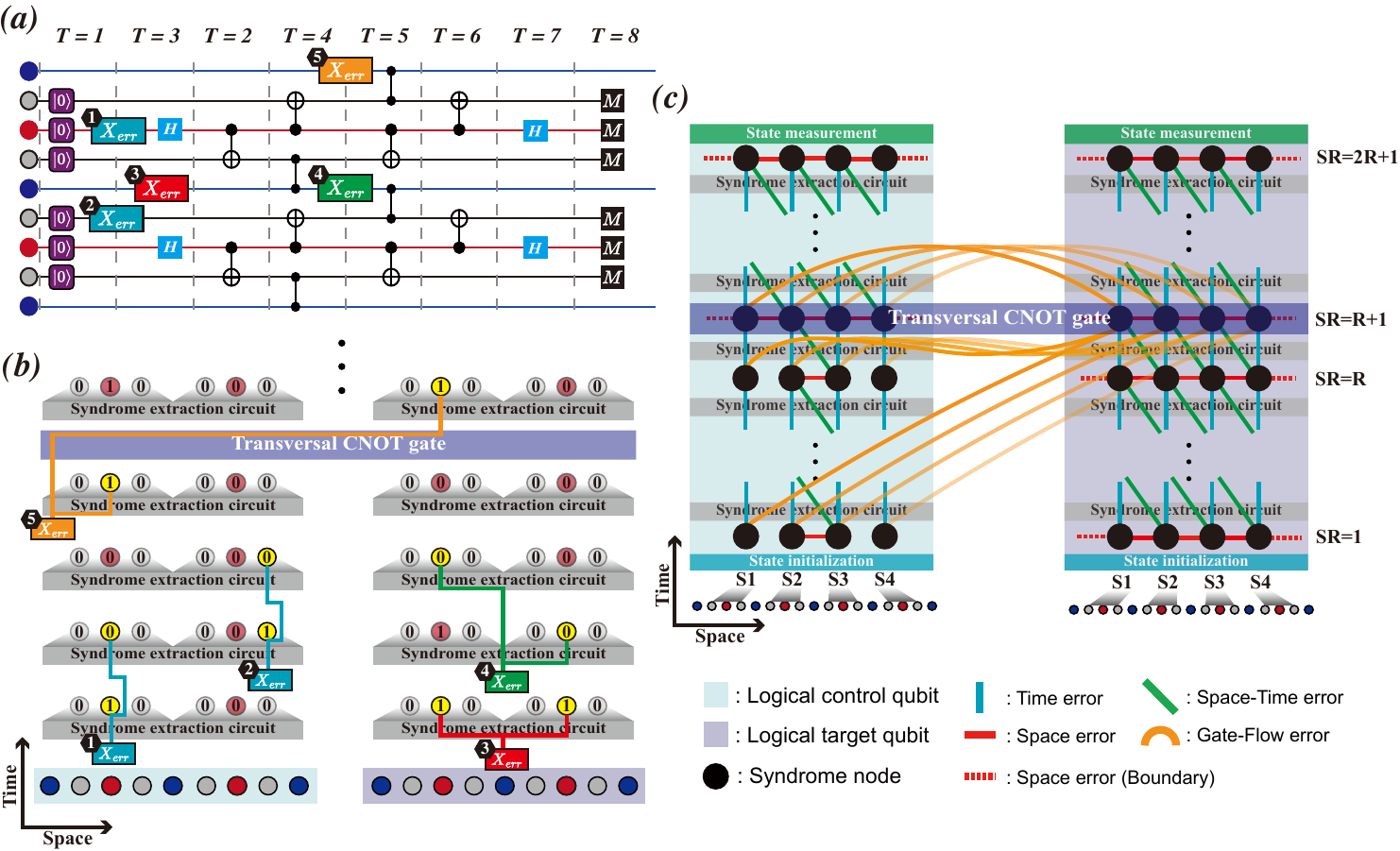}
    \caption{ Syndrome graph for two logical qubits considering the transversal CNOT gate. (a) Errors can be categorized into four types: Space (red box), Time (blue box), Space-Time (green box), and Gate-Flow errors (orange box). The numbers within the hexagon boxes are assigned to each of these errors: 1 and 2 for Time errors, 3 for a Space error, 4 for a Space-Time error, and 5 for a Gate-Flow error. These errors are depicted within a syndrome extraction circuit for a distance of 3. (b) An example showcases the outcomes of syndrome extraction circuits using 21 physical qubits, where an error can influence the outcomes of syndrome or flag qubits, affecting neighboring qubits horizontally, vertically, or diagonally. Additionally, a Gate-Flow error affects two syndrome qubits on both logical qubits. (c) The syndrome graph for the $d=5$ structure with \textbf{2R} rounds of syndrome extraction using Z stabilizers is depicted. The horizontal and vertical axes represent the space of the syndrome nodes (\textbf{S1}, \textbf{S2}, \textbf{S3}, and \textbf{S4}) and syndrome rounds (\textbf{SR}). The logical CNOT gate is executed at $\mathbf{SR=R+1}$, resulting in a total of $\mathbf{2R+1}$ syndrome rounds. Each logical qubit produces four syndrome nodes per syndrome round. Edges represent the correlations due to the four categorized errors (Space, Time, Space-Time, and Gate-Flow errors). A Space error on the boundary of each logical qubit connects with the boundary except on the logical control qubit before the transversal CNOT gate has been executed. }
    \label{fig3}
\end{figure*}

As we collect the data by running a quantum memory experiment on a quantum processor, we get the corresponding measured outcome of physical qubits based on errors that occurred as it is executed. The outcome can be mapped on the binary vector, a syndrome, considering space and time. The syndrome is calculated by checking the parity using the outcomes of each case sample as input. A detection event, resulting in a ``1" syndrome bit, occurs when the measurement outcomes of a syndrome extraction circuit involving flag and syndrome qubits differ from those of the previous extraction circuit and have odd parity \cite{repetition_flag}. 

An error that produces two detection events can be categorized into four types: Space, Time, Space-Time, and Gate-Flow errors. Any arbitrary error can be decomposed into a combination of these error categories \cite{correlated}. Figure \ref{fig3} (a) shows corresponding bit-flip errors in a syndrome extraction circuit using Z stabilizers, where blue, red, green, and orange boxes correspond to Space, Time, Space-Time, and Gate-Flow errors, respectively. The hexagon boxes with numbers correspond to 1 and 2 for Time errors, 3 for a Space error, 4 for a Space-Time error, and 5 for a Gate-Flow error in Figure \ref{fig3} (a) and (b). A Time error occurs when errors impact a syndrome or flag qubit, while other error types affecting data qubits differ based on when and where the error occurs within a subroutine of a quantum circuit.

Figure \ref{fig3} (b) illustrates how the measurement outcomes of flag and syndrome qubits change due to these categorized errors. The figure provides an example of outcomes from multiple rounds of syndrome extraction circuits, organized according to their execution time, with three data qubits for each logical qubit. A Space error affects neighboring syndrome qubits horizontally, while a Time error results from an error during the preparation of syndrome or flag qubits, causing outcome changes vertically. There is another type of error, the Space-Time error. The edge corresponds to an error that generates a pair of outcome changes diagonally during two-qubit gates in a syndrome extraction circuit on a data qubit. The case when an error spreads from one to the other logical qubit produces a Gate-Flow error. A Gate-Flow error produces outcome changes on syndrome qubits from both logical qubits. Two affected syndrome qubits due to a Gate-Flow error locally correspond to each logical qubit. 

We utilize a syndrome graph to decode errors and evaluate the performance of the transversal CNOT gate across multiple rounds of the syndrome extraction circuit. A syndrome graph is a 2D graph where each node and edge represents a syndrome bit or error, with the weight of each edge corresponding to the error probability \cite{topology_MWPM,subsystem_MWPM,XZZX_MWPM,pymatching}. The syndrome graph is constructed by considering the four categorized errors, enabling us to decode errors and calculate a correction operator. 

Figure \ref{fig3} (c) depicts the syndrome graph for a case where the syndrome extraction circuits are applied \textbf{2R} times, featuring a structure comprised of five data qubits for the Z basis of each logical qubit. The graph consists of two regions: a logical control and a target qubit. The horizontal and vertical axes represent space and time, respectively. During each syndrome round, every logical qubit generates four syndrome bits \textbf{S1}, \textbf{S2}, \textbf{S3}, and \textbf{S4}. Every syndrome bit is obtained by checking the parity with the measured outcomes of physical qubits while respecting their position and time within the code. The parameter \textbf{SR} implies a syndrome round in the graph obtained by comparing two consecutive results of syndrome extraction circuits. The syndrome extraction circuits used in the quantum memory experiment are depicted as grey boxes, showing their relative implementation timing concerning syndrome rounds. The transversal CNOT gate is implemented at the time $\mathbf{SR=R+1}$. Executing the syndrome extraction circuits for \textbf{R} rounds both before and after the transversal CNOT gate results in a total of $\mathbf{2R+1}$ syndrome rounds. These include the first and last temporal boundary, where we consider the state of initial and final measured data qubits. An edge represents an error that produces two detection events on the connected syndrome nodes by the edge. Each edge corresponds to one of the categorized errors in the syndrome graph. The weight of each edge in a syndrome graph corresponds to its error probability, calculated based on the error rates of individual single- and two-qubit gates according to the hardware specifications. These calculations are performed using the Stim code \cite{stim}.

\begin{figure*}[t]
    \centering
    \includegraphics[width=\textwidth]{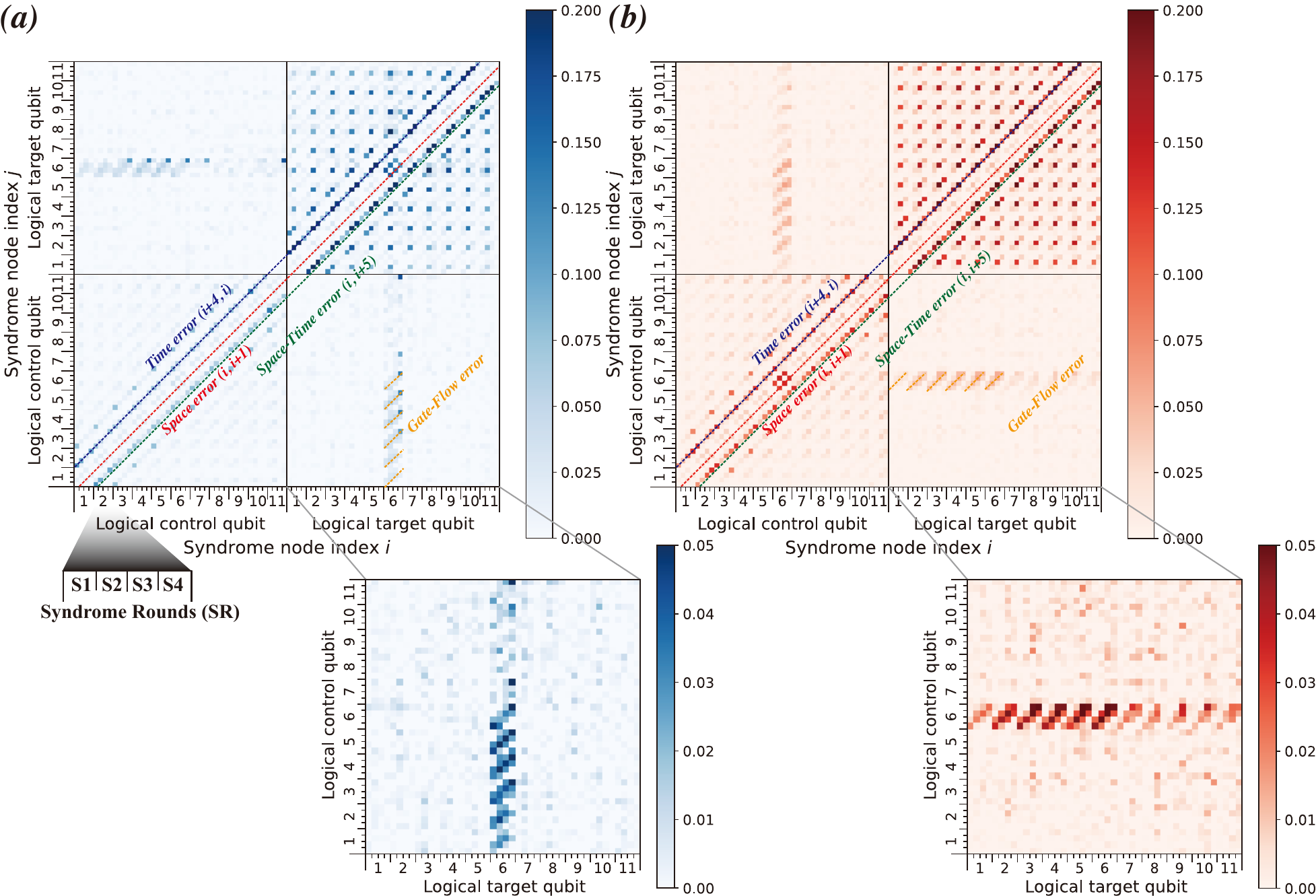}
    \caption{ Correlation matrices for the syndrome nodes (\texttt{ibm\_sherbrooke}). The two logical states, the logical control and target qubits, are initialized in the Z(X) basis, and each one has 5 data qubits. The data is obtained by sampling the quantum memory experiment $10^5$ times for both (a) $\ket{00}_L$ and (b) $\ket{++}_L$, with $\mathbf{2R=10}$ rounds of the syndrome extraction circuit. The axes represent the 11 blocks corresponding to syndrome rounds for a logical qubit, with each block displaying four syndrome nodes (\textbf{S1}, \textbf{S2}, \textbf{S3}, and \textbf{S4}) arranged according to their location within the code. A pixel represents the error probability that generates a ``1" syndrome bit on the corresponding i and j th syndrome nodes. The parameter $\mathbf{SR}$ corresponds to the syndrome round, and the transversal CNOT gate has been executed at $\mathbf{SR=6}$. The probability of Space, Time, Space-Time, and Gate-Flow errors are visually represented by red, blue, green, and orange dotted lines, respectively. The color bar is truncated to 0.2 and 0.05 to enhance the visibility. The same graph without the truncation can be seen in Appendix \ref{app:supp5.2}. }
    \label{fig4}
\end{figure*}

It is important to note that error propagation occurs only when an error happens within a time no later than the execution of the transversal CNOT gate. Therefore, error propagation occurs only when an error happens within a time no later than time $\mathbf{SR=R+1}$ corresponding to the execution of the transversal CNOT gate. As a result, all Gate-Flow errors connect syndrome nodes before $\mathbf{SR=R+1}$ on the logical control qubit and the syndrome nodes on the logical target qubit at $\mathbf{SR=R+1}$. A Space error on boundary data qubits of a logical qubit without error propagation produces only one detection event, necessitating a connection to the boundary \cite{MWPM,google_qec1,repetition_flag}. 

Logical qubits preserve data qubit information by detecting errors through the measured outcomes of syndrome and flag qubits from the syndrome extraction circuit and decoding them for error correction. We utilized the Minimum Weight Perfect Matching (MWPM) algorithm to rectify errors based on the outcomes obtained from the syndrome extraction circuits and the corresponding syndrome graph \cite{MWPM,MWPM_2,MWPM_3}. Given a sample obtained by conducting a quantum memory experiment on a device, we recovered the measured final data qubits using a correction operator that most likely cancels errors using the Pymatching algorithm \cite{pymatching,pymatching2}.

\section{Results}

The state of two logical qubits can be denoted as $\ket{\psi_C \psi_T}_L=\ket{\psi_C}_L \otimes \ket{\psi_T}_L $ where $\ket{\psi_C}_L$ and $\ket{\psi_T}_L$ represent the state of logical control and target qubits. When the logical qubits are in the Z basis, the state can be one of four computational basis states $\{ \ket{00}_L, \ket{01}_L, \ket{10}_L, \ket{11}_L \}$. Since a logical X operator can propagate from the logical control qubit to the logical target qubit through the transversal CNOT gate, these logical states should be transformed into $\{ \ket{00}_L, \ket{01}_L, \ket{11}_L, \ket{10}_L \}$ after passing the quantum memory experiment. In the case of the logical qubits in the X basis, a logical Z operator can be propagated reversely. The logical states $\{ \ket{++}_L, \ket{+-}_L, \ket{-+}_L, \ket{--}_L \}$ ideally change into $\{ \ket{++}_L, \ket{--}_L, \ket{-+}_L, \ket{+-}_L \}$ following the logical CNOT gate. We collect $10^5$ samples by conducting quantum memory experiments on a quantum device for each round and initial state with the \textit{qiskit} Python package \cite{qiskit}. A single round of syndrome extraction circuit took approximately $\sim1.9\mu$s and $\sim4.1\mu$s on \texttt{ibm\_torino} and \texttt{ibm\_sherbrooke}, respectively. 

Based on detection events, error patterns that include Gate-Flow errors can be seen through a correlation matrix. The correlation matrix can be plotted with the probability of errors as a function of the labels of syndrome nodes, $i$ and $j$, corresponding to an edge connecting the two syndrome nodes in the syndrome graph. Each error probability ($p_{ij}$) can be calculated as follows from the sampled data \cite{google_qec1}:

\begin{equation}
p_{ij}=\frac{1}{2} \left(1-\sqrt{1-\frac{4(\langle x_i x_j \rangle-\langle x_i \rangle\langle x_j \rangle)}{(1-2\langle x_i \rangle-2\langle x_j \rangle+4\langle x_i \rangle\langle x_j \rangle )}}\right)
\label{eq1}
\end{equation}

\noindent While $\langle x_i\rangle$ and $\langle x_j\rangle$ are the probability of the detection event at $i$ and $j$ th syndrome node, $\langle x_i x_j\rangle$ is the probability of two detection events at the same time. The correlation matrix is symmetric since $p_{ij}=p_{ji}$. We set negative values to zero to enhance visibility and clarity in the visualization.

Figure \ref{fig4} shows the correlation matrices for the case of the structure comprising 39 physical qubits and $\mathbf{2R=10}$ rounds of the syndrome extraction circuit both before and after the transversal CNOT gate. We show the correlation matrices obtained with data from \texttt{ibm\_sherbrooke}, while the equivalent figures using data from \texttt{ibm\_torino} employed 57 physical qubits are available in Appendix \ref{app:supp5.2}. The figure shows the correlation matrices for both the X and Z basis with labeled syndrome nodes. We label the number of syndrome nodes with priority given to space over time. We assign numbers first from the logical control qubit and then to the target logical qubit. Each correlation matrix can be divided into four regions depending on where $i$ and $j$ are: a pixel $(i,j)$ that can belong to 1) the logical control qubit, 2) the logical target qubit or 3) and 4) the correlation region where $i$ and $j$ are in different logical qubits. There are eleven syndrome rounds, with each round producing four syndrome bits for every logical qubit. The transversal CNOT gate is performed at $\mathbf{SR=6}$ following the five rounds of the syndrome extraction circuit. 

\begin{figure*}[ht]
    \centering
    \includegraphics[width=\textwidth]{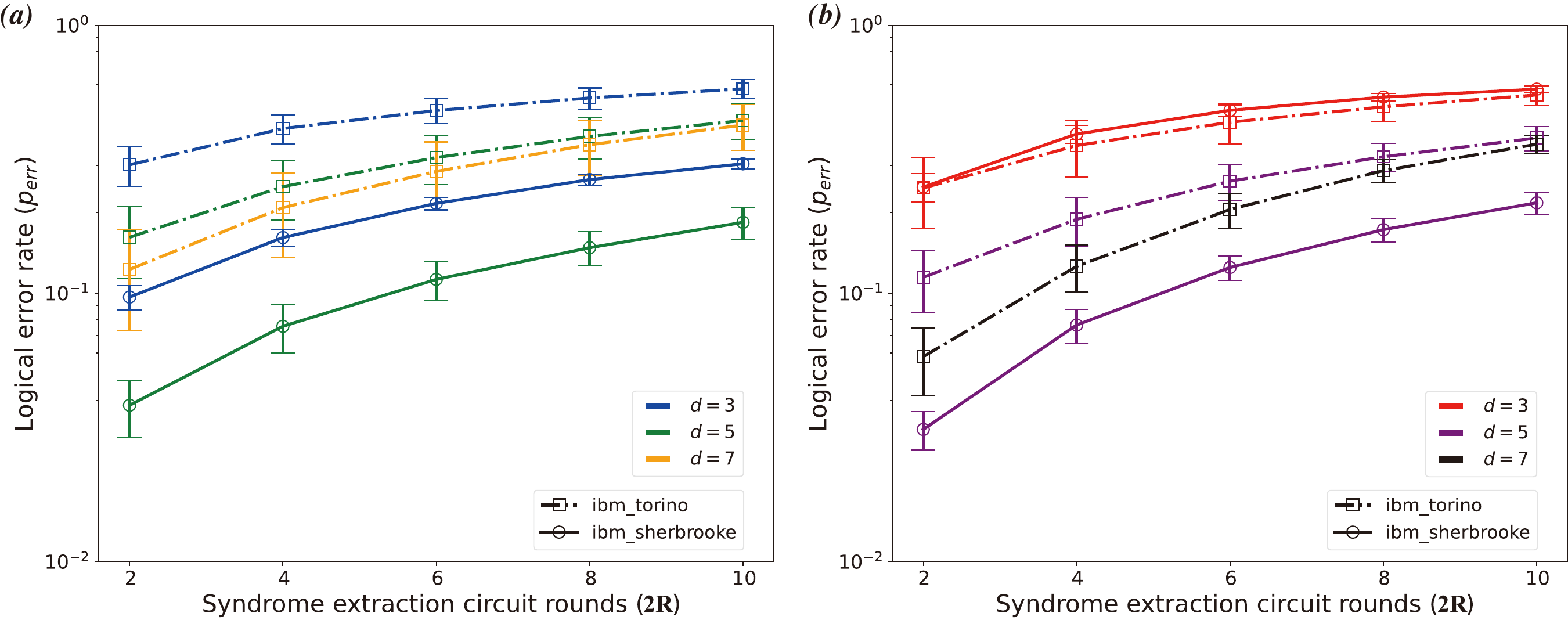}
    \caption{ Logical error rate suppression with code distance. The code considers (a) X errors using Z stabilizers or (b) Z errors using X stabilizers. The blue (red), green (purple), and orange (black) lines correspond to structure sizes of 21, 39, and 57 physical qubits correcting X (Z) errors, respectively. While solid lines denote the result from \texttt{ibm\_sherbrooke}, dash-dotted lines are those from \texttt{ibm\_torino}. The average logical error rates for four basis states are depicted with their standard deviation as a function of the number of syndrome extraction circuit rounds (\textbf{2R}), devices, and structure size ($d\in \{3,5,7\}$). Each data point is obtained from $10^5$ data samples collected from the corresponding devices. As the number of physical qubits increases, the average logical error rates decrease for both X and Z error cases. }
    \label{fig5}
\end{figure*}

There are spatial or temporal errors within each logical qubit region in the correlation matrix in Figure \ref{fig4} for the logical qubits in the (a) Z or (b) X basis. While measurement errors (Time error) correspond to the pixels at ($i+4, i$), data qubit errors (Space error) are represented by the pixels at ($i, i+1$). Space-Time errors can be seen in the pixels at ($i, i+5$). These errors are shown in the blue, red, and green diagonal dotted lines, respectively. While the probabilities of Space-Time errors are relatively lower than other types of errors, Time errors show the high error probabilities among the categorized errors for all cases. This matches the hardware specifications, where the dominant errors are readout errors.

The pixels representing Space errors at $\mathbf{SR=6}$ are notably highlighted with high error probabilities after executing the transversal CNOT gate. Based on the type of errors that are detected by logical qubits, the highly visible values are located at either the logical control or target qubit. For example, the error probabilities of data qubits in the logical target qubit at $\mathbf{SR=6}$ are relatively high, in Figure \ref{fig4} (a), where the logical qubits consider bit-flip errors. Conversely, regarding the Z errors in Figure \ref{fig4} (b), spatial errors in the logical control qubit at $\mathbf{SR=6}$ occur more frequently than at other times and spaces. From the perspective of the affected logical qubit, it is because additional errors are introduced not only from the other logical qubit through the transversal CNOT gate but also as it executes the gate including ancilla qubits. 

A Gate-Flow error can be seen in the area within a correlation between the logical control and target qubit. The Gate-Flow errors are visible in the correlation matrix in distinct ways, depending on the type of error, whether they are bit- or phase-flip errors. Gate-Flow errors indeed exhibit conspicuous error probabilities, particularly at specific syndrome rounds. Notably, in Figure \ref{fig4}, at $\mathbf{SR=6}$, we observe this phenomenon prominently in the (a) logical target or (b) control qubit. Diagonal lines represent the dominance of Gate-Flow errors. These diagonal pixels represent the two syndrome nodes located at the same positions in each logical qubit, exhibiting strong correlation mainly due to error propagation while respecting spatial constraints. Additionally, notable correlations exist with locally neighboring syndrome nodes. This phenomenon suggests that the execution of the transversal CNOT gate can result in Gate-Flow errors, causing temporal correlations between syndrome nodes. In Ref. \cite{ion_cnot}, the work also demonstrates strong correlations with the stabilizers of two logical qubit blocks due to error propagation. However, the key difference lies in the number of syndrome extraction circuits. We not only show strong correlations due to Gate-Flow errors but also highlight the existence of long temporal correlations between syndrome nodes by executing multiple rounds of a syndrome extraction circuit before the logical CNOT gate.

To evaluate the performance of the transversal CNOT gate, we compute the logical error rate ($p_{err}$), referred to as infidelity. This is determined by the expectation values of the logical Pauli operators of two logical qubits for the error-corrected final states, following the methodology described in Refs. \cite{trapped_ion_transversal}. We consider the \textbf{2R} rounds of the syndrome extraction circuits (\textbf{R} rounds before and \textbf{R} rounds after the logical CNOT gate) and correct errors using the decoder. We obtain logical error rates for structures with distances of 3 and 5, requiring 21 and 39 physical qubits on \texttt{ibm\_sherbrooke}, respectively. Despite the device's structure allowing for a mapping distance of 7, the presence of broken edges between physical qubits in a specific region, which means that two physical qubit gates have error rate as 1, makes it challenging to execute quantum circuits, as shown in Appendix \ref{app:supp6}. Additionally, we evaluate the performance of the transversal CNOT gate on \texttt{ibm\_torino} for structures comprising 3, 5, and 7 data qubits per logical qubit. Experiments are conducted for each round, considering initial states for both the Z and X bases. Logical error rates are evaluated for each device based on collected samples.

Figure \ref{fig5} shows the logical error rates averaged across all four states of either the X or Z basis, along with their standard deviation. The logical error rates of the transversal CNOT gate are plotted as a function of the structure size and the number of syndrome extraction circuits for each device. The results demonstrate a reduction in the average logical error probabilities as the code size increases when the transversal CNOT gate is executed with syndrome extraction circuits. This trend remains consistent across all rounds of syndrome extraction circuits. Notably, among the selected physical qubits for experiments, the average error rates of two-qubit gates on \texttt{ibm\_torino} are higher than those on \texttt{ibm\_sherbrooke}, whereas the average readout error rates on \texttt{ibm\_sherbrooke} are higher than on \texttt{ibm\_torino}. More details can be found in Appendix \ref{app:supp5.1}. Furthermore, according to the correlation matrices in Appendix \ref{app:supp5.2}, \texttt{ibm\_torino} has experienced more correlated errors, which inherently occur as the real device executes a quantum circuit, such as leakage errors \cite{leakage_err}. Given that readout and correlated errors are the predominant source of errors, these device properties may contribute to higher logical error rates on \texttt{ibm\_torino} compared to \texttt{ibm\_sherbrooke}. 

The gap in logical error rates between different structure sizes diminishes as the number of rounds of a syndrome extraction circuit increases. Notably, this phenomenon is particularly pronounced in the results from \texttt{ibm\_torino}. This trend could be attributed to the increased rounds of syndrome extraction per logical CNOT gate, which may introduce greater ambiguity in the correction operator calculated by the decoder owing to Gate-Flow errors \cite{decoding_cnot}.

\section{CONCLUSIONS AND DISCUSSIONS}
The scalability of fault-tolerant quantum computers constructed with Stabilizer codes relies on the potential of error suppression achieved by increasing the size of their code block. This property must be demonstrated in both logical qubits and gates during error correction. However, demonstrating error suppression of a logical CNOT gate with multiple rounds of error correction remains an open question in the NISQ era.

In this study, we evaluated the performance of the logical CNOT gate with multi-cycle syndrome extraction circuits both before and after the gate ($\mathbf{\{R=1,2,3,4,5\}}$), for a total $\mathbf{2R}$ rounds, and compared their effectiveness across different sizes ($d\in\{3, 5, 7\}$). We prepared logical qubits using the Repetition code with flag qubits and implemented the logical CNOT gate using physical CNOT gates transversely, leveraging auxiliary qubits. We conducted experiments on the IBM quantum machines \texttt{ibm\_sherbrooke} and \texttt{ibm\_torino}. The results show that despite the emergence of Gate-Flow errors caused by the transversal CNOT gate, which creates long temporal correlations, increasing the number of physical qubits on IBM quantum processors suppresses errors. Furthermore, error suppression persists during multiple rounds of the syndrome extraction circuit. Additionally, we also performed experiments on \texttt{ibm\_brisbane}, another IBM hardware platform. We observed lower logical error rates as we increased the code size from three data qubits to seven per logical qubit by correcting bit-flip errors. However, in contrast, when correcting phase-flip errors, the logical error rates with seven data qubits were higher than those with three and five data qubits. Further details can be seen in Appendix \ref{app:supp5.4}.

The post-selection method has been used as a key technique in recent experimental demonstrations of quantum error correction in the NISQ era \cite{ion_cnot,entangling_gate,ibm_transversal,google_qec1,carbon,heavy-hex-entangle,teleportation,ibm_qec1,ibm_qec2}. This method demonstrates evaluating the performance of stabilizer codes by discarding any samples with non-trivial errors or selecting non-trivial samples that might be correctable. However, in this work, we calculated the logical error rates without using the post-selection method. Hence, this work can be expanded by incorporating the post-selection method, which could potentially lower logical error rates.

It's important to note that the Repetition code is unable to rectify both types of errors (bit- and phase-flips) simultaneously. Furthermore, additional flag qubits in the Repetition code may lower the effectiveness of the distance which is closely related to the number of correctable errors \cite{repetition_flag}. However, this study primarily focused on investigating the transversal CNOT gate with varying structure sizes, detecting errors multiple times, and demonstrating error suppression through increased code size. Therefore, the study showed that the transversal CNOT gate can be executed while correcting one type of error on real quantum hardware.

While the transversal CNOT gate has been designed to cause fewer Gate-Flow errors by only interacting locally between corresponding data qubits of two logical qubits, these errors can still negatively impact logical qubits during execution on quantum hardware. This issue has also been discussed in references \cite{decoding_cnot,ion_cnot,heavy-hex-entangle}. 

Ref. \cite{decoding_cnot} addresses the challenge of reducing detrimental effects by employing a decoder that utilizes a hypergraph for decoding complex errors resulting in more than two detection events. Furthermore, they investigated how the performance of a series of logical gates changes with varying numbers of syndrome extraction circuits per logical gate. A promising avenue for future work would involve comparing the performance of the logical CNOT gate on real quantum devices using a correlated decoder versus the MWPM decoder, examining how this performance varies with the number of error detection cycles to find the optimized number of cycles. It would be enlightening to see whether a correlated decoder can also handle more complex Gate-Flow errors, which move through logical gates over time. Another line of investigation would be to train and apply machine learning based decoders which have recently shown promising performance \cite{ANN_Dec, ANN_IBM,google_RNN}.

Finally, an open question for future research involves observing Gate-Flow errors occurring simultaneously in both directions between logical qubits. Notably, a logical Bell state has been successfully constructed using quantum hardware across different platforms \cite{ion_cnot,carbon,heavy-hex-entangle}. Therefore, a promising approach is to construct a logical Bell state and calculate detection event probabilities for each syndrome bit across multiple cycles of syndrome rounds. It would be intriguing to observe how the detection event probabilities of X and Z stabilizers change before and after the execution of the transversal CNOT gate and whether the existence of temporal correlations occurred by Gate-Flow errors. 

\section*{Data Availability}
The data that support the findings of this study can be provided upon reasonable request to the corresponding author.

\section*{Code Availability}
The source code used to generate figures in this work can be provided upon reasonable request to the corresponding author.

\section*{ACKNOWLEDGMENT}
YK acknowledges the support of the CSIRO Research Training Program Scholarship and a University of Melbourne Research Training Scholarship. The University of Melbourne supported the research through the establishment of the IBM Quantum Network Hub at the University.

\section*{Competing Financial Interests}
The authors declare no competing financial or non-financial interests.

\clearpage
\begin{table*}[htbp]
\caption{ List the hardware specifications used in the experiments. The values are obtained from the calibration table, where 21 physical qubits are used with $\mathbf{2R=10}$ rounds of syndrome extraction circuits in the Z basis. The table displays the type of two-qubit gates used in each hardware, including ECR pulse, CZ, or CX gates. It also depicts the time length required for the execution of a single-qubit(SQ), two-qubit(TQ) gate, and measurement. Furthermore, it shows the median error rate for each gate. The experiments on each device were conducted in the duration listed in the table. }

\resizebox{\textwidth}{!}{
\begin{tabular}{ |c | c | c | c | c | c | c | c | c| }
\hline
\multirow{2}{*}{Hardware} & \multirow{2}{*}{TQ Gate Type} & \multicolumn{3}{c|}{Gate Time (ns)} & \multicolumn{3}{c|}{Median Error Rate (Median)} & Experiment Duration\\ 
 & & SQ Gate & TQ Gate & Measurement & SQ Gate & TQ Gate & Measurement & (YYYY-MM-DD, 00:00:00+00:00)\\ \hline

\multirow{2}{*}{\texttt{ibm\_sherbrooke}} & \multirow{2}{*}{ECR} & \multirow{2}{*}{56.88} & \multirow{2}{*}{533.33} & \multirow{2}{*}{1244.44} & \multirow{2}{*}{$2.27\times 10^{-4}$} & \multirow{2}{*}{$7.72\times 10^{-3}$} & \multirow{2}{*}{$1.10\times 10^{-2}$} & 2024-03-15\\
 &  &  &  &  &  &  &  & 06:20:36 $\sim$ 11:30:28 \\ \hline
 
\multirow{2}{*}{\texttt{ibm\_brisbane}} & \multirow{2}{*}{ECR} & \multirow{2}{*}{60} & \multirow{2}{*}{660} & \multirow{2}{*}{4,000} & \multirow{2}{*}{$2.29\times 10^{-4}$} & \multirow{2}{*}{$8.02\times 10^{-3}$} & \multirow{2}{*}{$1.33\times 10^{-2}$} & 2024-04-23 \\
 &  &  &  &  &  &  &  & 01:48:20 $\sim$ 03:51:17 \\ \hline
 
\multirow{2}{*}{\texttt{ibm\_torino}} & \multirow{2}{*}{CZ} & \multirow{2}{*}{32} & \multirow{2}{*}{84} & \multirow{2}{*}{1,560} & \multirow{2}{*}{$2.78\times 10^{-4}$} & \multirow{2}{*}{$4.78\times 10^{-3}$} & \multirow{2}{*}{$1.93\times 10^{-2}$} & 2024-04-25  \\
 &  &  &  &  &  &  & & 03:38:43 $\sim$ 06:23:11 \\ \hline
 
\texttt{FakeWashington} & \multirow{2}{*}{CX} & \multirow{2}{*}{35.55} & \multirow{2}{*}{469.33} & \multirow{2}{*}{864} & \multirow{2}{*}{$2.81\times 10^{-4}$} & \multirow{2}{*}{$1.14\times 10^{-2}$} & \multirow{2}{*}{$1.47\times 10^{-2}$} & \multirow{2}{*}{-} \\
(Classical Simulator) &  &  &  &  &  &  & & \\ \hline
 
\end{tabular}}
\label{table2}
\end{table*}

\appendix
\section{Quantum hardware \& Classical simulator}\label{app:supp1}
The experiments were conducted using four different hardware platforms, including processors covered in the main paper. These platforms are \texttt{ibm\_sherbrooke}, \texttt{ibm\_brisbane}, \texttt{ibm\_torino}, and the \texttt{Fakewashington} simulator. The \texttt{Fakewashington} simulator is a classical simulator that performs all circuits with the \textit{qiskit} Python package using the ``fake provider" module in fake (simulated) backend classes \cite{qiskit}. 

In these hardware setups, quantum gates are executed using a basis gate set that ensures universality. IBM's hardware uses the following basis gates to execute arbitrary quantum circuits: $X$, $\sqrt{X}$, $R_Z$ (single-qubit gates), and a two-qubit gate such as CZ (Controlled-Z) or CNOT (Controlled-NOT). Additionally, some platforms use specialized gates like the ECR pulse (echoed cross-resonance gate) to achieve gate universality and create entanglement between pairs of physical qubits \cite{ECR}. For example, \texttt{ibm\_sherbrooke} and \texttt{ibm\_brisbane} use the ECR pulse as their basis gate, enabling the construction of physical CNOT and CZ gates using single-qubit and ECR gates. Conversely, \texttt{ibm\_torino} and the \texttt{Fakewashington} simulator use CZ and CX gates, respectively, to achieve quantum gate universality.

The hardware specifications, including the relaxation time (T1) and coherence time (T2) of each physical qubit, are listed in a calibration table. This table also provides error rates for both single- and two-qubit gates, as well as for qubit measurements. All physical qubit gates, except for the virtual Z gate ($R_Z$), require a specific execution time, which is detailed in the calibration table. Periodically, this calibration table is updated with new values, except for the \texttt{Fakewashington} simulator. Each dataset in this work is analyzed using the corresponding device's calibration table to assess logical error rates, taking into consideration the completion time of each experiment. 

Table \ref{table2} presents the implementation duration for experiments on each hardware platform, along with gate times and median error rates for physical gates. The calibration table is derived from experiments utilizing 21 physical qubits with $\mathbf{2R=10}$ rounds of a syndrome extraction circuit. For the \texttt{Fakewashington} simulator, we utilized the most recently updated values available in the calibration table. Among real quantum hardware platforms, \texttt{ibm\_torino} stood out with the shortest gate time and the lowest median error rate for its two-qubit gates; however, it had the highest median readout errors. In contrast, \texttt{ibm\_sherbrooke} demonstrated the shortest time required and the lowest median readout errors for qubit measurements, while \texttt{ibm\_brisbane} required the longest time for qubit measurements.

\section{Quantum circuit optimization}\label{app:supp2}
We utilized the Python library $qiskit$ and transpiled circuits to execute the quantum circuits on the device. To optimize the quantum circuits and obtain results, at the hardware level, we employed the following methods:

\begin{itemize}
    \item Dynamic Decoupling method (XXXX): The method involves allowing free evolution of data qubits while syndrome and flag qubits are being measured. However, this can lead to idling errors for the data qubits. To mitigate this, we apply the Dynamic Decoupling method to all data qubits, minimizing their idling error \cite{DD}.
    \item Conditional reset gate: After each round of the syndrome extraction circuit, flag and syndrome qubits are initialized to $\ket{0}$. To efficiently achieve this, we use conditional reset gates that incorporate conditional bit-flip operations based on the measured outcomes immediately after qubit measurements, reducing the resetting time \cite{ibm_qec2}.
    \item ALAP scheduling: We utilize the ``As Late As Possible" (ALAP) scheduling method to optimize the timing of physical qubit gates at the hardware level, ensuring efficient execution of quantum circuits \cite{qiskit}.
\end{itemize}

\section{Sampling process}\label{app:supp3}
A quantum memory experiment is designed to evaluate the transversal CNOT gate, incorporating either X or Z syndrome extraction circuits based on the logical qubit basis. During each experiment run, data were sampled by collecting outcomes from the syndrome and flag qubits for each round of the syndrome extraction circuit, including the final states of the data qubits. We conducted sampling $10^5$ times for all cases, encompassing different distances, numbers of syndrome extraction circuits, and initial states. However, when conducting experiments on the \texttt{Fakewashington} simulator, we sampled $2 \times 10^4$ times.

\subsection{X stabilizers}\label{app:supp3.1}
When the state of the logical qubit is in the Z basis, we employ Z stabilizers, as detailed in the main paper. However, we can also consider the state of the logical qubit in the X basis by utilizing the phase-flip code with flag qubits to correct Z errors. The stabilizers for the X basis are defined using the X Pauli operators of neighboring data qubits, represented as $X_i X_{i+1}$, where $i$ denotes the data qubit number. For example, when a logical qubit consists of 3 data qubits, it can be protected from phase-flip errors and defined in the logical states ($\ket{+}_L=\ket{+++}$, $\ket{-}_L=\ket{---}$). The logical Pauli operators for the logical qubit are selected similarly to when using Z stabilizers. Single- and two-qubit gates are organized in the same sequence as for Z stabilizers, with the exception that the two-qubit gates between data qubits and their neighboring flag qubits are physical CNOT gates instead of CZ gates. Hadamard gates are utilized for initializing data qubits or measuring their final state in the Z basis.

\subsection{Quantum memory experiment}\label{app:supp3.2}
\begin{figure*}[htbp]
  \centering
  \includegraphics[width=\textwidth]{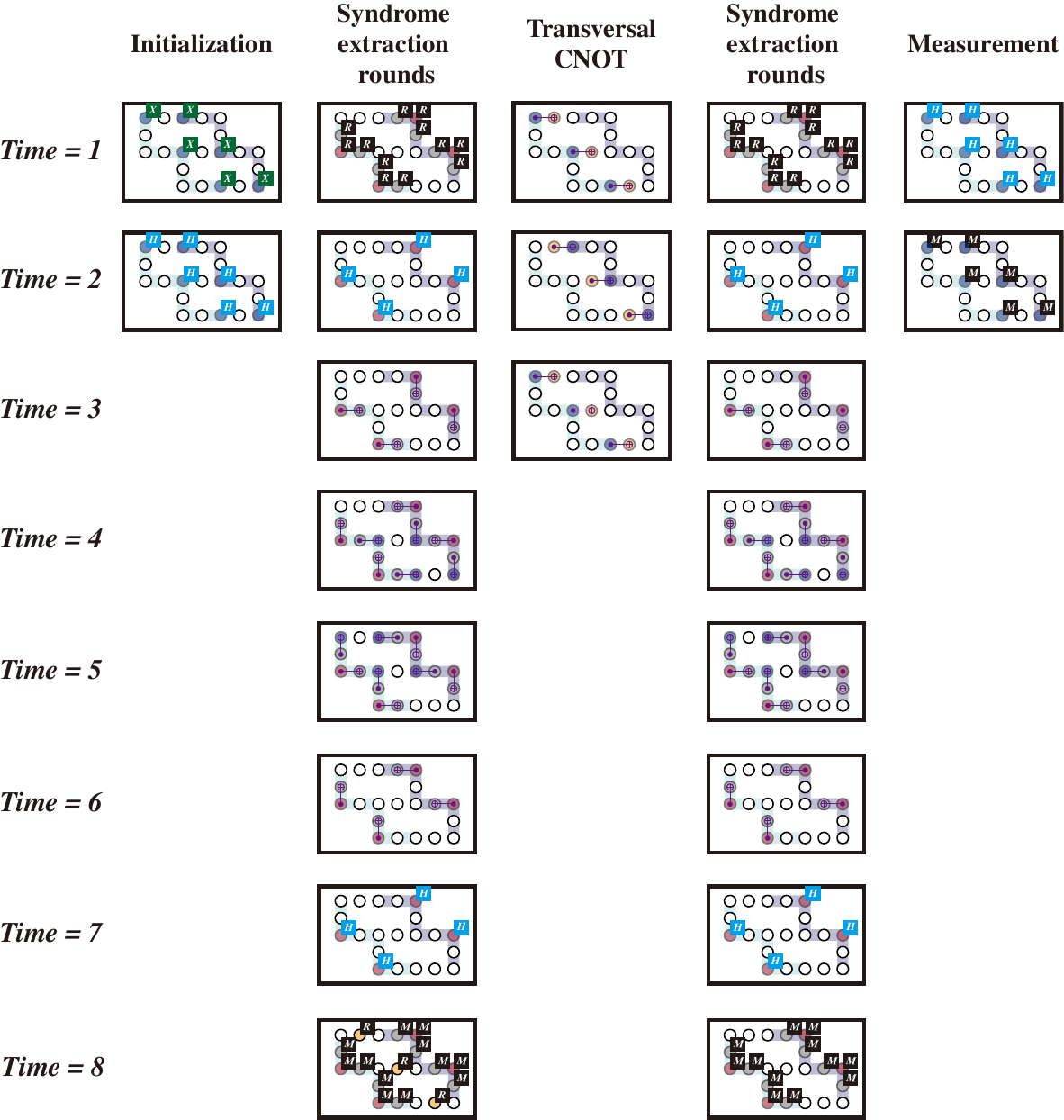}
  \caption{ Quantum memory experiment for sampling the $\ket{--}_L$ with 21 physical qubits. Each node represents the physical qubits and according to their type, the color of the node is blue, red, orange, and grey, corresponding data, syndrome, ancilla, and flag qubits. The quantum circuit has five steps: Initialization, Syndrome extraction rounds, Transversal CNOT, and Measurement.  The horizontal line represents each step, and the vertical line corresponds to the sequence of gates and time. When physical qubits are inactive and undergoing free evolution the color of the node is filled with white. }
  \label{fig6}
\end{figure*}

Figure \ref{fig6} depicts the quantum circuit for sampling the data with $\ket{--}_L$ structure with one round ($\mathbf{R=1}$) of the syndrome extraction circuit considering X stabilizers, before and after the logical CNOT gate. The circuit can be divided into five steps as follows:
\begin{enumerate}
    \item Initialization: The logical control qubit, which comprises the three data qubits, and the other is the logical target qubit, consisting of the rest of the data qubits. Every data qubit is initialized as $\ket{-}$ by applying X and Hadamard gates. Hence, two logical qubits are initialized as $\ket{--}_L$. 
    \item Syndrome extraction rounds: Data qubits undergo a syndrome extraction circuit considering X stabilizers. At the last round of the syndrome extraction circuits before the transversal CNOT gate, we initialize ancilla qubits while measuring flag and syndrome qubits.
    \item Transversal CNOT: The transversal CNOT gate is applied to two logical qubits. Physical CNOT gates are employed between a data qubit from the logical control qubit and its closest auxiliary qubit in the structure. The physical CNOT gates are applied to grouped physical qubits to implement the transversal CNOT gate, as described in the main paper.
    \item Syndrome extraction rounds: Another round of the syndrome extraction circuit is executed to measure syndrome and flag qubits for tracking errors in the system.
    \item Measurement: Finally, we measure all data qubits in the Z basis using Hadamard gates before the measurement.
\end{enumerate}

\subsection{Syndrome}\label{app:supp3.3}
Execution of noisy syndrome extraction circuits results in a binary vector determined by their outcomes. In the absence of errors, both syndrome and flag measurements are deterministic, hence the product of measurement outcomes is also deterministic \cite{stim}. A syndrome is designed to be a zero vector when no errors are present in the system. The observation of a 1 bit in the syndrome indicates the presence of an error, and these 1 bits enable a decoder to identify the most likely correction operator based on their locations. Depending on the stabilizer group, bit- or phase-flip errors involving flag qubits result in a 1 bit in the syndrome. The Repetition code with Z stabilizers can detect X errors, resulting in 1 bits in the syndrome, but it cannot detect Z errors. Conversely, the code with X stabilizers can track Z errors. We employ the same parity check process described in Ref. \cite{repetition_flag} for two logical qubits in the structure, producing the corresponding syndrome to rectify errors.

\clearpage
\section{Decoding process}\label{app:supp4}
\begin{figure*}[htbp]
  \centering
  \includegraphics[width=\textwidth]{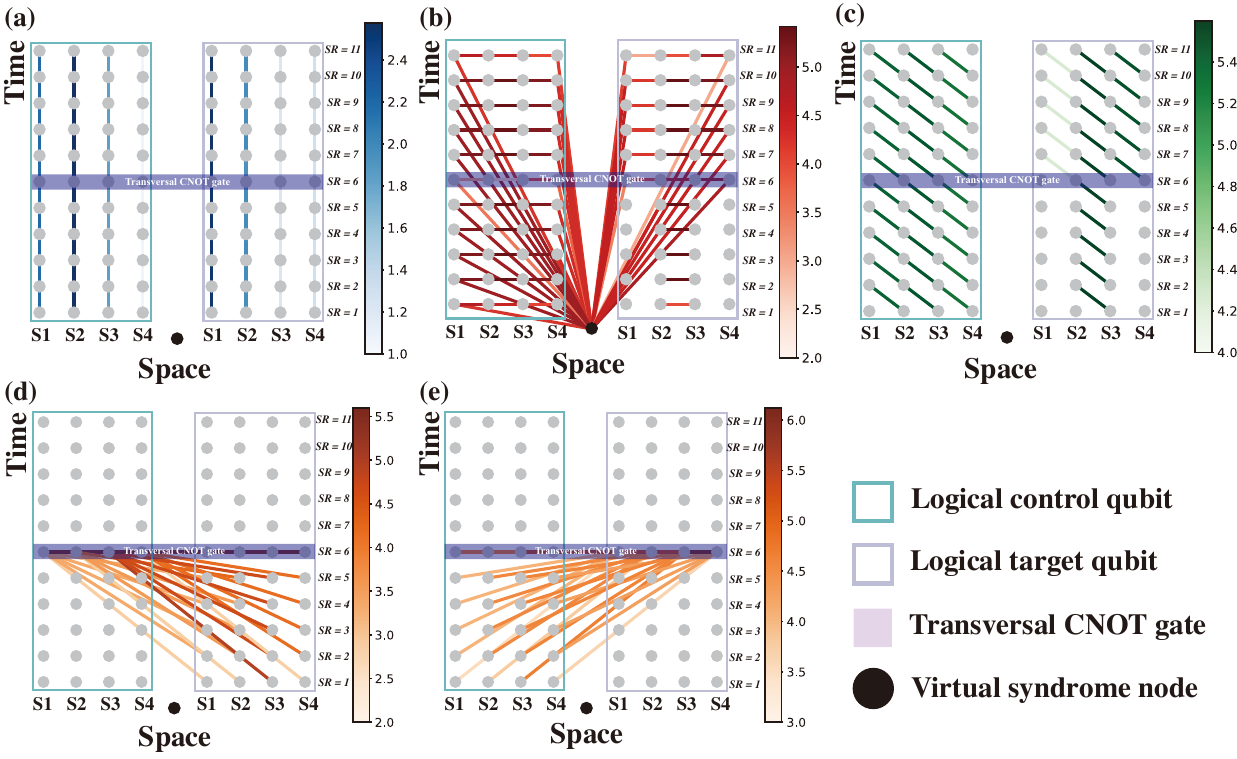}
  \caption{ Syndrome graphs with the types of edges for the structure with 39 physical qubits. The logical control and target qubits are enclosed with the blue and purple empty boxes. The axes represent the space of the syndrome bit and the time of the syndrome round (\textbf{SR}). The figure is the case where the syndrome extraction is implemented five times, producing eleven syndrome rounds. The transversal CNOT gate is implemented at $\mathbf{SR=6}$ and depicted as the solid purple box. Every node corresponds to the syndrome bit arranged by their location and time. Edges represent (a) Time, (b) Space, (c) Space-Time, and (d, e) Gate-Flow errors and they have their weights corresponding to their error probabilities. The black dot represents the virtual syndrome node, used to display the S errors on the boundary of the logical qubit. The decoder considers Gate-Flow errors connecting one syndrome node, located on the logical (d) target or (e) control qubit before the transversal CNOT operation, to a syndrome positioned on the other logical qubit at $\mathbf{SR=6}$. These scenarios correspond to cases where the logical qubits are initialized in either the X or Z basis states. }
  \label{fig7}
\end{figure*}

\subsection{Circuit level noise model}\label{app:supp4.1}
Each physical qubit gate is susceptible to errors. The noisy gate can be decomposed into its ideal gate and error channels. We employ a uniform depolarizing error channel simplified with Pauli operators. When the error rate $p$ is derived from the calibration table for a basis gate, we implement the circuit-level noise model using the depolarizing error model as follows:

\begin{itemize}
  \item The uniform Pauli error channel, characterized by $p_X$, $p_Y$, and $p_Z$, satisfies the condition $p_X = p_Y = p_Z = p/3$ and occurs following an ideal single-qubit gate.
  \item The uniform Pauli error channel is applied to two qubits, excluding the identity operator ($I\otimes I$), following the ideal two-qubit gates (CNOT). The probability of each Pauli error from $\{X,Y,Z,I\}^{\otimes2}/\{I\otimes I\}$ is $p/15$. Additionally, two-qubit depolarization error channels are individually employed before and after the two-qubit gates, respecting their corresponding single-qubit gate error rates in the calibration table \cite{repetition_flag}. When a device uses an ECR pulse as a basic gate, we use the same error channel for a CZ gate as a physical CNOT gate. However, if the device can operate a CZ pulse at the hardware level, we only consider the two-qubit depolarization error channel.
  \item An X error channel with the probability of $p$ is introduced before the measurement and after the reset gate.
  \item When a physical qubit is inactive and undergoes free evolution, it encounters the uniform Pauli error channel for a single qubit with the probability of the error being $p$, the idling error rate from the calibration table.
\end{itemize}

We adopt the above circuit level noise model to calculate weights for each edge and construct a syndrome graph using Stim code \cite{stim}, the syndrome graph is used for decoding errors using the Pymatching \cite{pymatching,pymatching2}. 

\subsection{Syndrome graph}\label{app:supp4.2}

Figure \ref{fig7} displays the syndrome graphs for the structure with five data qubits and eleven syndrome rounds, where the horizontal and vertical axes represent space and time. The graph can be divided into two regions: the logical control and target qubits. The region for the logical control qubit is enclosed within the empty blue box, while that of the other logical qubit is enclosed within the empty purple box. Each logical qubit has four syndrome nodes (\textbf{S1}, \textbf{S2}, \textbf{S3}, and \textbf{S4}) per syndrome round (\textbf{SR}). The syndrome graphs are the same, but the only difference lies in the type of edges. Edges in the graph correspond to four types (Space,  Time,  Space-Time, and Gate-Flow errors) of errors that produce a pair of corresponding syndrome nodes to have detection events. We depict the categorized errors in \ref{fig7} (a), (b), (c), and (d) for the logical qubits that are initialized as $\ket{++}_L$. Notably, the correlated edges can differ in two ways depending on what error type we consider. Hence, the Gate-Flow errors for the Z basis ($\ket{00}_L$) are also depicted in a different way to those of the X basis in Figure \ref{fig7} (e). This is because when the logical qubits are in the X basis, a Z error can propagate from the logical target qubit to the logical control qubit and it is the opposite for the Z basis with a X error.

A virtual syndrome node is introduced to consider an error that produces an odd number of detection events, especially, in the case of a Space error on the boundary data qubit of the logical control qubit in Figure \ref{fig7} (b). However, when a Space error occurs before the transversal CNOT gate on the boundary data qubit of the logical target qubit, the virtual node is unnecessary. This is because the error will propagate to the other logical qubit and produce another detection event resulting in an even number of detection events being considered as a Gate-Flow error. The locally corresponding syndrome nodes are connected by a Gate-Flow error. Gate-Flow errors are only connected with syndrome nodes on the logical (d) target or (e) control qubit at the time $\mathbf{SR=6}$ when the logical gate has been executed depending on the basis of the logical qubits. Since errors propagate only when the transversal CNOT gate is implemented,  no Gate-Flow errors are considered after the gate.  This is attributed to the Pauli operators propagation rule which has been demonstrated in the main paper.

Each edge is assigned a weight based on the probability of the corresponding error, with a lower weight indicating a higher probability. When calculating a correction operator based on the syndrome, edges with lower weights are more likely to be chosen as correction operators to rectify errors. The figures are generated by considering the calibration table and using the Stim code considering the circuit-level noise. The weights of Time errors have relatively lower values than those of other errors. In Figure \ref{fig7}, these weights are calculated and displayed based on the calibration table of \texttt{ibm\_sherbrooke}. Depending on what physical qubit we use, the higher the measurement error rates in the hardware specification, the lower the weight. The weights of Gate-Flow errors exhibit the largest gap between the highest and lowest values among the error types. The edges in the first syndrome round have relatively lower values compared to other syndrome rounds. This could be because not only the data qubit's idling error and errors during the syndrome extraction circuit but also initialization errors on the data qubit affect it.

\clearpage
\section{Results}\label{app:supp5}
\subsection{Selected physical qubits}\label{app:supp5.1}

\begin{table*}[htb]
  \resizebox{\textwidth}{!}{
    \begin{tabular}{||c | c | c | c | c | c||} 
     \hline
     Qubit & T1($\mu$s) & T2($\mu$s) & Readout Error Rate& SQ Gate Error Rate& TQ Gate Error Rate\\ 
     \hline\hline
    0 & 242.56 & 282.67 & 0.0109 & 0.00015 & [0.0053, 0.0078] \\ 
     \hline 
    1 & 339.82 & 195.64 & 0.0092 & 0.00014 & [0.0053, 0.0058] \\ 
     \hline 
    2 & 61.36 & 74.59 & 0.0071 & 0.00054 & [0.0162, 0.0058, 0.0061] \\ 
     \hline 
    3 & 349.3 & 274.39 & 0.0132 & 0.00014 & [0.0058, 0.0061] \\ 
     \hline 
    4 & 472.45 & 124.04 & 0.0159 & 0.00018 & [0.0058, 0.0071, 0.0107] \\ 
     \hline 
    5 & 120.51 & 130.06 & 0.0081 & 0.00036 & [0.0102, 0.0107] \\ 
     \hline 
    6 & 313.28 & 207.58 & 0.009 & 0.00014 & [0.007, 0.0102, 0.0077] \\ 
     \hline 
    7 & 185.15 & 183.28 & 0.0053 & 0.00017 & [0.006, 0.0077] \\ 
     \hline 
    8 & 437.75 & 359.17 & 0.0071 & 0.00019 & [0.006, 0.0085, 0.0074] \\ 
     \hline 
    9 & 90.06 & 34.51 & 0.01167 & 0.00035 & [0.0078, 0.008] \\ 
     \hline 
    10 & 208.48 & 60.85 & 0.0119 & 0.00019 & [0.0071, 0.0082] \\ 
     \hline 
    11 & 268.68 & 125.12 & 0.0511 & 0.00052 & [0.0085, 0.0077] \\ 
     \hline 
    12 & 149.35 & 267.64 & 0.02833 & 0.00021 & [0.008, 0.008] \\ 
     \hline 
    13 & 218.14 & 30.84 & 0.0094 & 0.00043 & [0.0101, 0.008] \\ 
     \hline 
    14 & 419.31 & 254.69 & 0.0073 & 0.00013 & [0.0101, 0.0057, 0.0088] \\ 
     \hline 
    15 & 381.73 & 174.45 & 0.0175 & 0.00015 & [0.0057, 0.0079] \\ 
     \hline 
    16 & 192.67 & 56.09 & 0.0278 & 0.00031 & [0.0082, 0.0079, 0.008] \\ 
     \hline 
    17 & 386.96 & 13.18 & 0.0568 & 0.01038 & [0.008, 0.0051] \\ 
     \hline 
    18 & 292.84 & 95.83 & 0.0048 & 0.00027 & [0.0041, 0.0084, 0.0051] \\ 
     \hline 
    19 & 242.84 & 300.14 & 0.0222 & 0.00021 & [0.0041, 0.0052] \\ 
     \hline 
    20 & 178.56 & 95.54 & 0.0142 & 0.00017 & [0.0077, 0.0096, 0.0052] \\ 
     \hline 
    \end{tabular}
  }\caption{ Calibration table values for 21 physical qubits used for experiments (\texttt{ibm\_sherbrooke}). The values obtained from the calibration table correspond to the experiment for $|00\rangle_L$ with ten rounds. The table lists the properties of physical qubits for the $d=3$ structure selected within the hardware. These include T1, and T2 time, as well as error rates of readout, single- (SQ) and two-qubit (TQ) gates. More than one two-qubit gate can be implemented and their error rates are listed accordingly. }
  \label{table_sherbrooke}
\end{table*}

Table \ref{table_sherbrooke} presents details such as the number of physical qubits, T1, T2, and error rates for gates, including single- and two-qubit gates in the case of \texttt{ibm\_sherbrooke}. These details are relevant to our quantum memory experiment utilizing $\ket{00}_L$ with 21 physical qubits. The average values of T1 and T2 are $264.4\mu$s and $159.1\mu$s, respectively. The average error rate for measurement gates is $0.017$, while for single-qubit gates it is $0.00073$, and for two-qubit gates, it is $0.0072$. The same vaules for other devices are available in Appendix \ref{app:supp6}.

Figure \ref{fig8} illustrates the specifications of the selected physical qubits for the $\ket{00}_L$ structure with distances 3 and 5 on \texttt{ibm\_sherbrooke}. The same figures for other devices can be seen in Appendix \ref{app:supp6}. Each node and edge in the figure represents a physical qubit and its connectivity with neighboring physical qubits, indicating possible direct two-qubit gate operations. The color of each node and edge corresponds to its readout error rate and two-qubit gate error rate, with brighter regions indicating higher error occurrence. Additionally, the average values of T1 and T2 among the selected physical qubits are depicted on the subplot corresponding to the 0th physical qubit. These values are represented as black lines on the left and right axes, with red and blue bars indicating T1 and T2 of the physical qubit, respectively. The background of the subplots displays the error rate of the single-qubit gate. Subplots for other physical qubits also show their T1, T2, and error rate of the single-qubit gate. It's observed that when more physical qubits are used, error rates for readout tend to increase. Furthermore, while the average error rates of two-qubit gates on \texttt{ibm\_torino} were 0.0066 and 0.0095 for distances 3 and 5, respectively, those on \texttt{ibm\_sherbrooke} were higher at 0.0072 and 0.011. However, the average readout error rates on \texttt{ibm\_sherbrooke} were 0.017 and 0.02, which are lower than those on \texttt{ibm\_torino} at 0.02 and 0.024. 

\begin{figure*}[t]
    \centering
    \includegraphics[width=\textwidth]{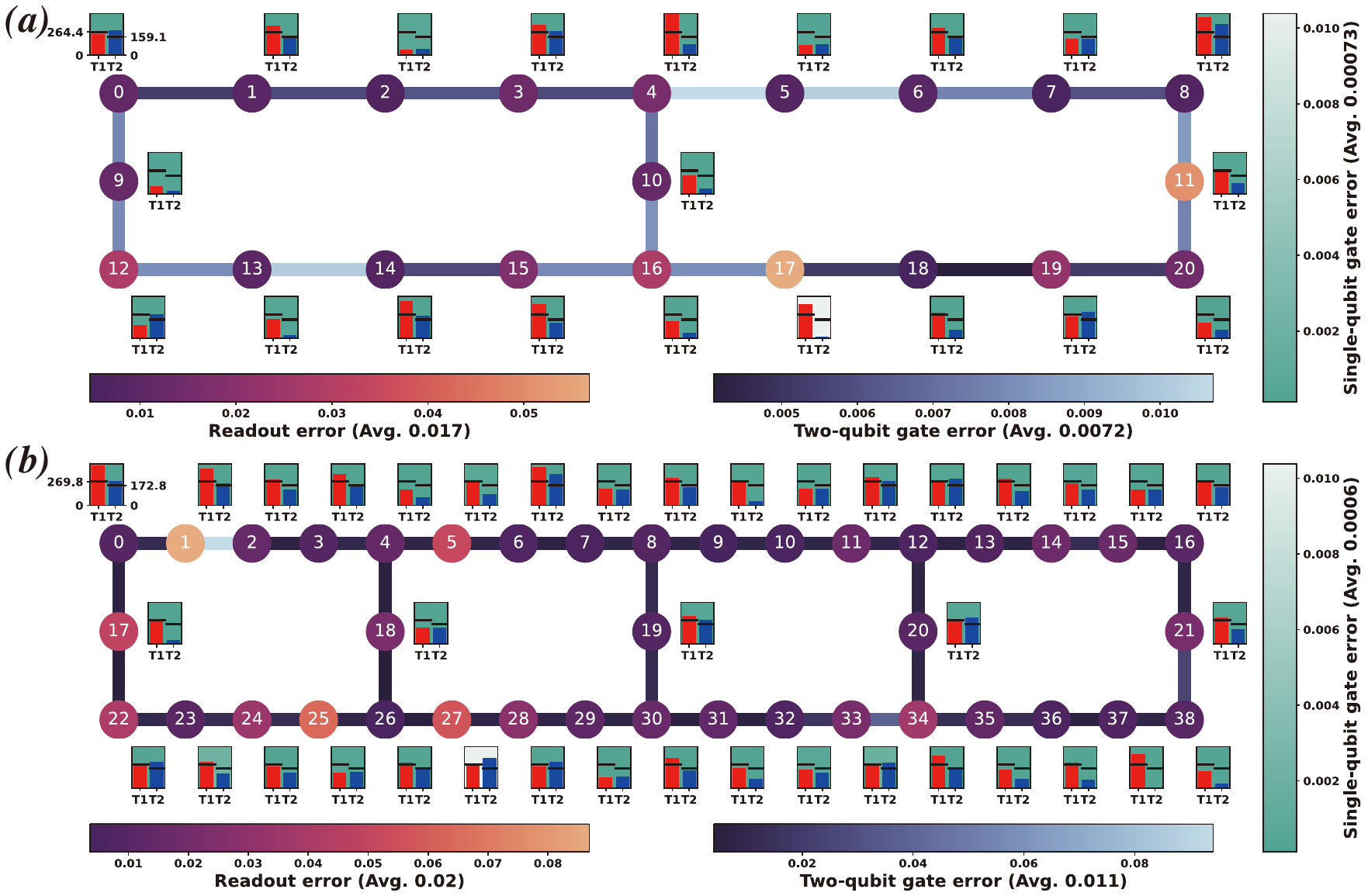}
    \caption{ Hardware specifications of physical qubits with colormap and bar charts (\texttt{ibm\_sherbrooke}). The graph with subplots shows the physical qubits' specification of the structure in (a) $d=3$ and (b) $d=5$, with $\ket{00}_L$ state. Each node and edge correspond to the physical qubit and connectivity of a two-qubit gate (ECR). The error rates of readout and two-qubit gate are displayed with colors. The subplot next to every node shows T1 and T2 times corresponding to their physical qubit. The average time of T1 and T2 are plotted as the black lines on the left and right sides. Their values can be seen on the subplot of the 0th physical qubit. The background of a subplot is colored corresponding to the error rate of the single-qubit gate. }
    \label{fig8}
\end{figure*}

\clearpage
\subsection{Correlation matrices}\label{app:supp5.2}

\begin{figure*}[t]
    \centering
    \includegraphics[width=\textwidth]{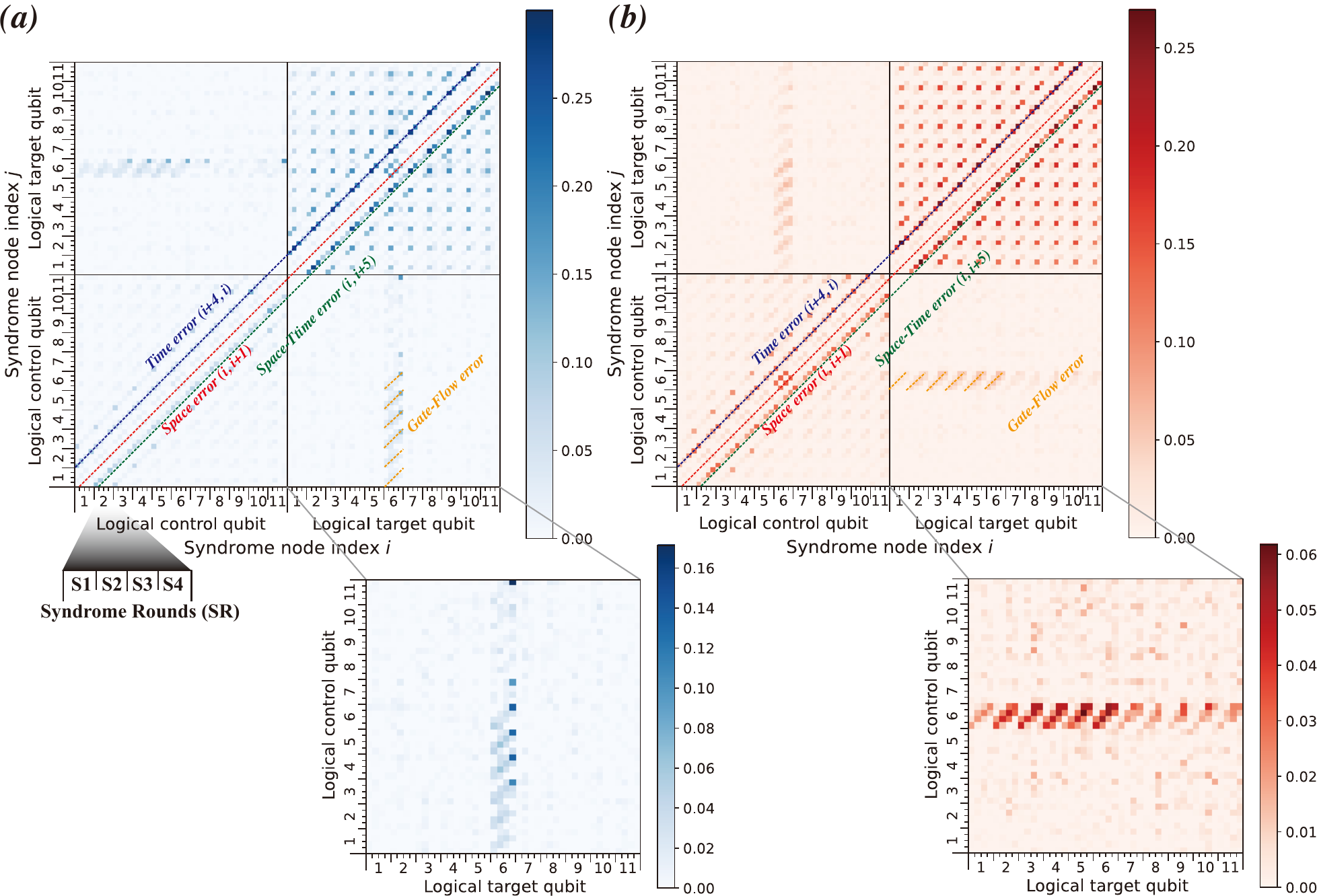}
    \caption{ Correlation matrices for the syndrome nodes (\texttt{ibm\_sherbrooke}). The two logical states, the logical control and target qubits, are initialized in the Z(X) basis, and each one has seven data qubits. The data is obtained by sampling the quantum memory experiment $10^5$ times for both (a) $\ket{00}_L$ and (b) $\ket{++}_L$, with $\mathbf{2R=10}$ rounds of the syndrome extraction circuit. The axes represent the eleven blocks corresponding to syndrome rounds for a logical qubit, with each block displaying six syndrome nodes arranged according to their location within the code. A pixel represents the error probability that generates a ``1" syndrome bit on the corresponding $i$ and $j$ th syndrome nodes. The parameter \textbf{SR} corresponds to the round of detection event, and the transversal CNOT gate has been executed at $\mathbf{SR=6}$. The probability of Space, Time, Space-Time, and Gate-Flow errors are visually represented by red, blue, green, and orange dotted lines, respectively. }
    \label{fig9}
\end{figure*}

\begin{figure*}[t]
    \centering
    \includegraphics[width=\textwidth]{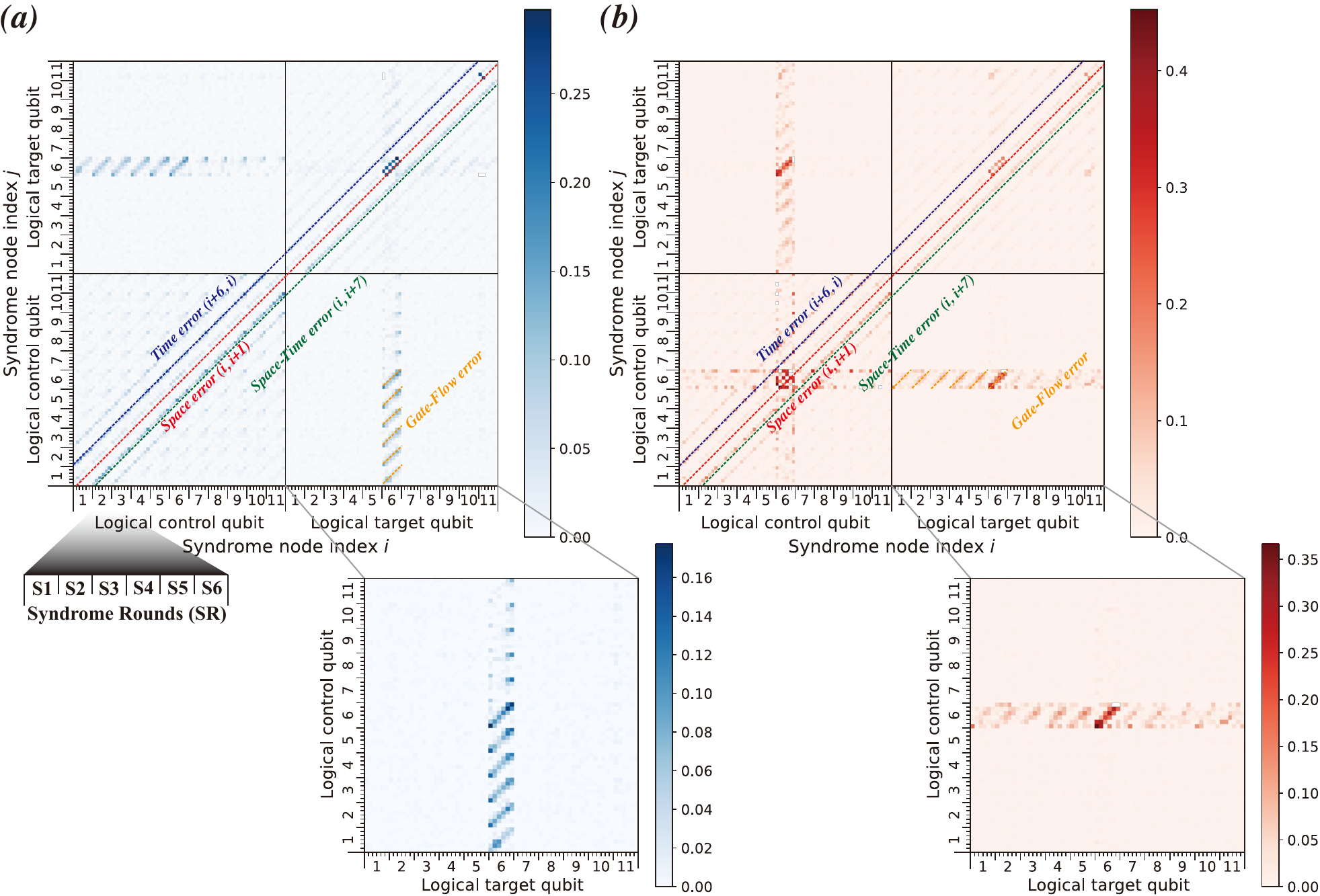}
    \caption{ Correlation matrices for the syndrome nodes (\texttt{ibm\_brisbane}). The two logical states, the logical control and target qubits, are initialized in the Z(X) basis, and each one has seven data qubits. The data is obtained by sampling the quantum memory experiment $10^5$ times for both (a) $\ket{11}_L$ and (b) $\ket{--}_L$, with $\mathbf{2R=10}$ rounds of the syndrome extraction circuit. The axes represent the eleven blocks corresponding to syndrome rounds for a logical qubit, with each block displaying six syndrome nodes arranged according to their location within the code. A pixel represents the error probability that generates a ``1" syndrome bit on the corresponding $i$ and $j$ th syndrome nodes. The parameter \textbf{SR} corresponds to the round of detection event, and the transversal CNOT gate has been executed at $\mathbf{SR=6}$. The probability of Space, Time, Space-Time, and Gate-Flow errors are visually represented by red, blue, green, and orange dotted lines, respectively. }
    \label{fig10}
\end{figure*}

\begin{figure*}[t]
    \centering
    \includegraphics[width=\textwidth]{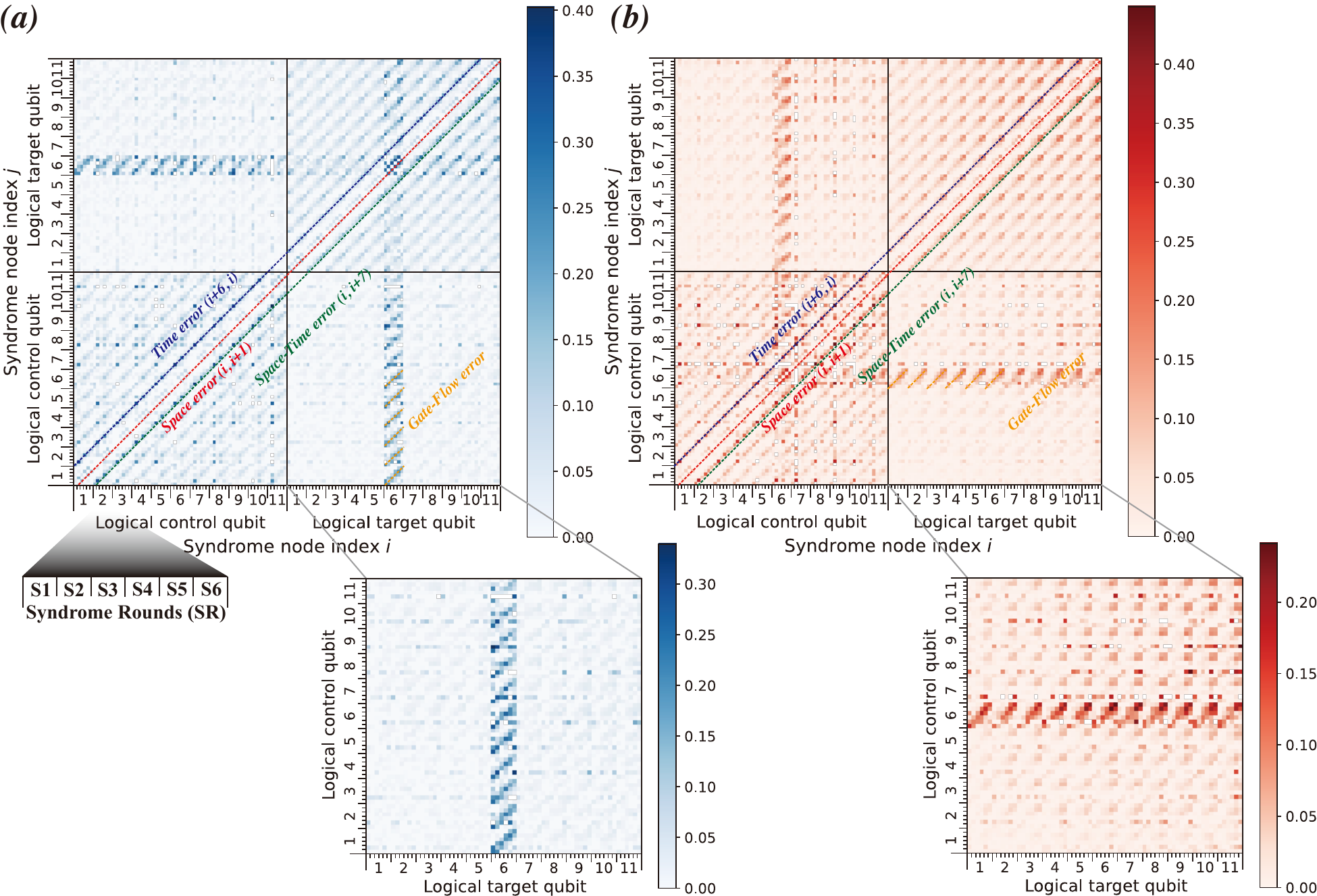}
    \caption{ Correlation matrices for the syndrome nodes (\texttt{ibm\_torino}). The two logical states, the logical control and target qubits, are initialized in the Z(X) basis, and each one has seven data qubits. The data is obtained by sampling the quantum memory experiment $10^5$ times for both (a) $\ket{11}_L$ and (b) $\ket{--}_L$, with $\mathbf{2R=10}$ rounds of the syndrome extraction circuit. The axes represent the eleven blocks corresponding to syndrome rounds for a logical qubit, with each block displaying six syndrome nodes arranged according to their location within the code. A pixel represents the error probability that generates a ``1" syndrome bit on the corresponding $i$ and $j$ th syndrome nodes. The parameter \textbf{SR} corresponds to the round of detection event, and the transversal CNOT gate has been executed at $\mathbf{SR=6}$. The probability of Space, Time, Space-Time, and Gate-Flow errors are visually represented by red, blue, green, and orange dotted lines, respectively. }
    \label{fig11}
\end{figure*}

\begin{figure*}[t]
    \centering
    \includegraphics[width=\textwidth]{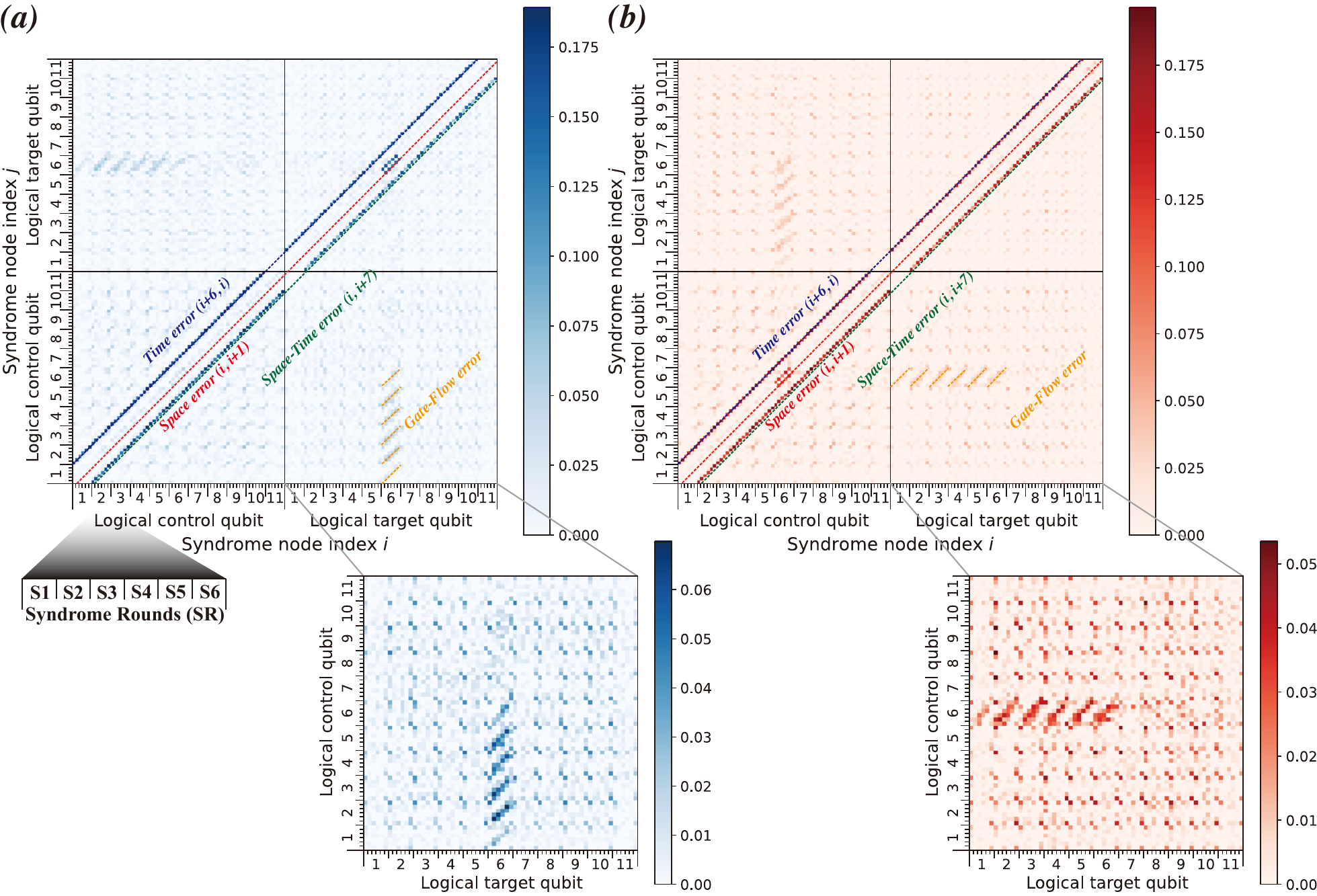}
    \caption{ Correlation matrices for the syndrome nodes (\texttt{Fakewashington}). The two logical states, the logical control and target qubits, are initialized in the Z(X) basis, and each one has seven data qubits. The data is obtained by sampling the quantum memory experiment $2 \times10^4$ times for both (a) $\ket{11}_L$ and (b) $\ket{--}_L$, with $\mathbf{2R=10}$ rounds of the syndrome extraction circuit. The axes represent the eleven blocks corresponding to syndrome rounds for a logical qubit, with each block displaying six syndrome nodes arranged according to their location within the code. A pixel represents the error probability that generates a ``1" syndrome bit on the corresponding $i$ and $j$ th syndrome nodes. The parameter \textbf{SR} corresponds to the round of detection event, and the transversal CNOT gate has been executed at $\mathbf{SR=6}$. The probability of Space, Time, Space-Time, and Gate-Flow errors are visually represented by red, blue, green, and orange dotted lines, respectively. }
    \label{fig12}
\end{figure*}

There are two methods to label the number of syndrome nodes: 1) labeling the nodes based on space first, and 2) labeling the nodes based on time first instead of space. We consider labeling the syndrome nodes for the logical control qubit first, followed by the logical target qubit in Figure \ref{fig4} of the main paper. Figure \ref{fig9} shows the same data as that without truncation. Figure \ref{fig10}, \ref{fig11}, and \ref{fig12} present correlation matrices derived from sampled data obtained from different hardware platforms \texttt{ibm\_brisbane}, \texttt{ibm\_torino}, and \texttt{Fakewashington}, respectively. Each matrix corresponds to experiments conducted with initial states (a) $\ket{11}_L$ and (b) $\ket{--}_L$. In this setup, a logical qubit comprises seven data qubits, necessitating a total of 57 physical qubits for the structure with ten rounds of a syndrome extraction circuit (\textbf{2R}). They show high error probabilities, particularly for Time errors. Moreover, the probabilities of Space errors on either the logical control or target qubit are significantly influenced by the basis of the logical qubits. X(Z) errors on the logical target (control) qubit before the logical CNOT gate propagate to the other logical qubit, resulting in Gate-Flow errors, which can be seen in the figures. Strong correlations are observed at specific times, notably at $\mathbf{SR=6}$, between two logical qubits from the perspective of the affected logical qubit due to Gate-Flow errors. Overall, these correlation matrices demonstrate similar patterns of correlations as observed in results from both actual devices and the classical simulator. It is worth pointing out that in the results from \texttt{ibm\_brisbane}, the correlation matrix with the X basis shows high error probabilities in the correlation region at $\mathbf{SR=6}$ for both logical qubits. It could be related to the biased errors during the execution of the logical CNOT gate. Additionally, it is important to note that although the \texttt{Fakewashington} simulator tends to have higher average error rates for readout and two-qubit gates compared to real devices, the highest error probabilities within the correlation matrices from the real devices are approximately twice as large as those from the classical simulator.

\begin{figure*}[t]
    \centering
    \includegraphics[width=\textwidth]{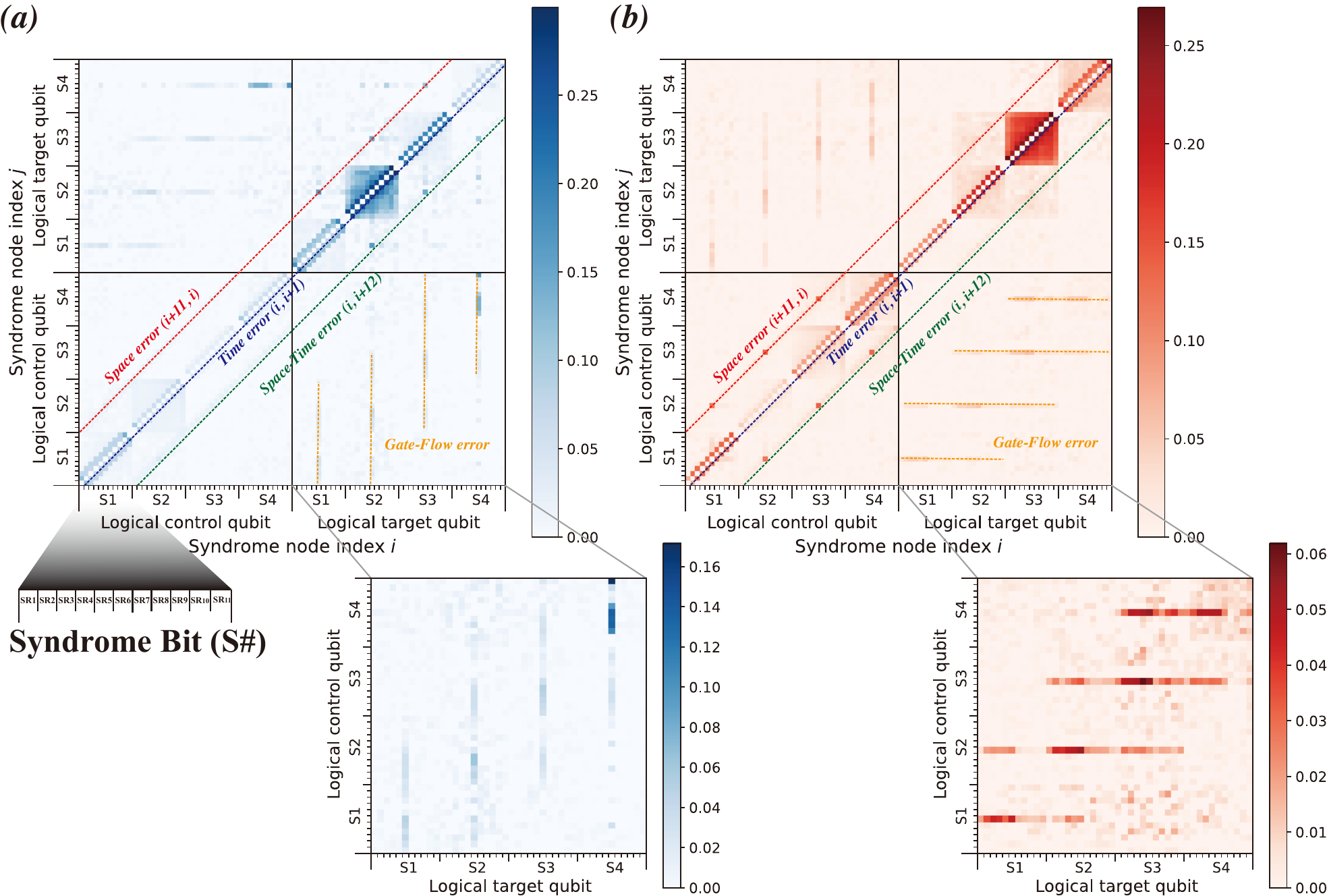}
    \caption{ Correlation matrices for the syndrome nodes (\texttt{ibm\_sherbrooke}). The syndrome nodes are labeled based on time before space. The two logical states, the logical control and target qubits, are initialized in the Z(X) basis, and each one has five data qubits. The data is obtained by sampling the quantum memory experiment $10^5$ times for both (a) $\ket{00}_L$ and (b) $\ket{++}_L$, with $\mathbf{2R=10}$ rounds of the syndrome extraction circuit. The axes represent the four blocks corresponding to syndrome nodes for a logical qubit, with each block displaying eleven syndrome rounds arranged according to their location within the code. A pixel represents the error probability that generates a ``1" syndrome bit on the corresponding $i$ and $j$ th syndrome nodes. The syndrome nodes are labeled first based on time, and then on space. The parameter \textbf{SR} corresponds to the round of detection event, and the transversal CNOT gate has been executed at $\mathbf{SR=6}$. The probability of Space, Time, Space-Time, and Gate-Flow errors are visually represented by red, blue, green, and orange dotted lines, respectively. }
    \label{fig13}
\end{figure*}

\begin{figure*}[t]
    \centering
    \includegraphics[width=\textwidth]{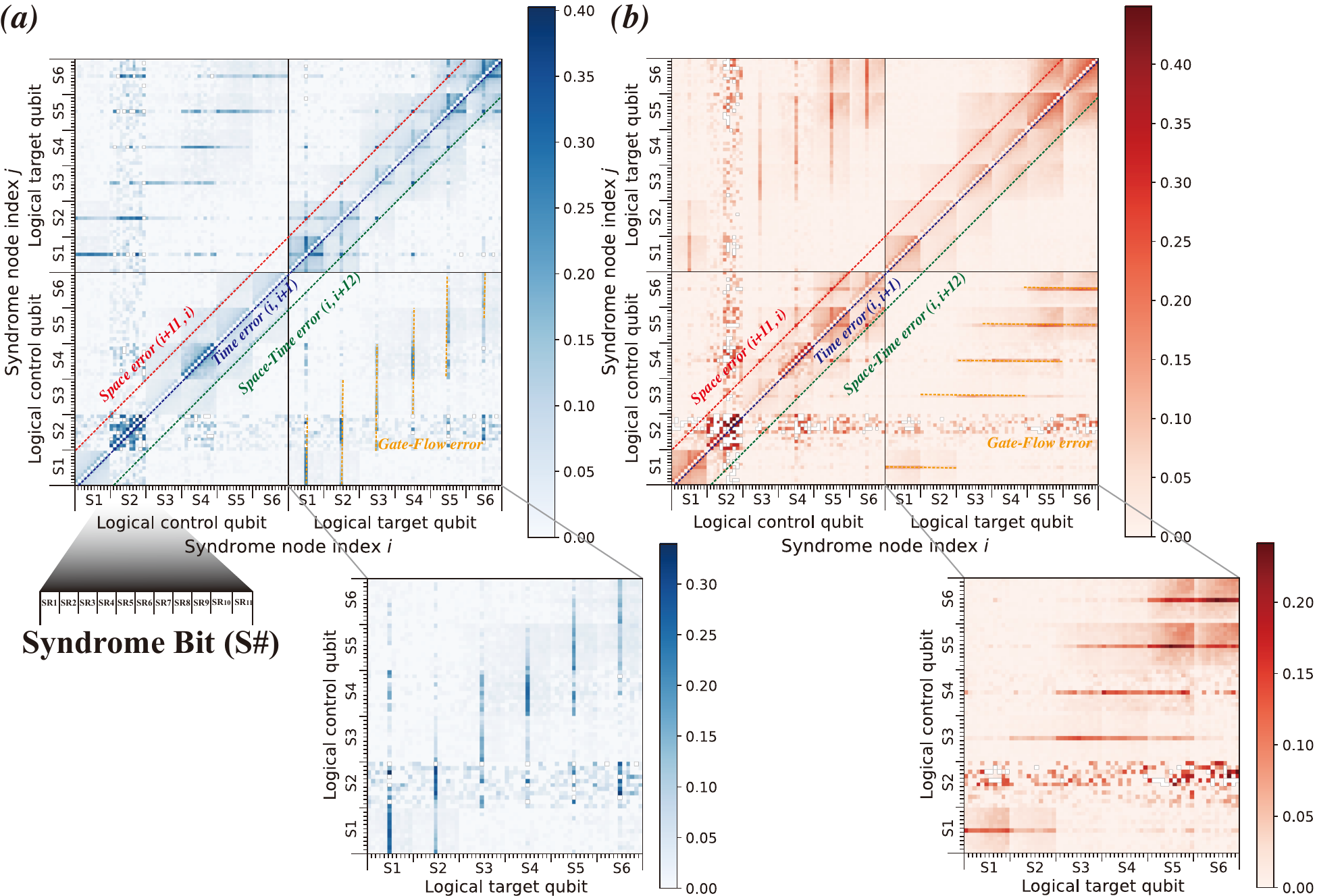}
    \caption{ Correlation matrices for the syndrome nodes (\texttt{ibm\_torino}). The syndrome nodes are labeled based on time before space. The two logical states, the logical control and target qubits, are initialized in the Z(X) basis, and each one has seven data qubits. The data is obtained by sampling the quantum memory experiment $10^5$ times for both (a) $\ket{11}_L$ and (b) $\ket{--}_L$, with $\mathbf{2R=10}$ rounds of the syndrome extraction circuit. The axes represent the six blocks corresponding to syndrome nodes for a logical qubit, with each block displaying eleven syndrome rounds arranged according to their location within the code. A pixel represents the error probability that generates a ``1" syndrome bit on the corresponding $i$ and $j$ th syndrome nodes. The syndrome nodes are labeled first based on time, and then on space. The parameter \textbf{SR} corresponds to the round of detection event, and the transversal CNOT gate has been executed at $\mathbf{SR=6}$. The probability of Space, Time, Space-Time, and Gate-Flow errors are visually represented by red, blue, green, and orange dotted lines, respectively. }
    \label{fig14}
\end{figure*}

Figures \ref{fig13} and \ref{fig14} illustrate correlation matrices obtained by labeling the syndrome nodes based on time before space in the axes of the correlation matrices from \texttt{ibm\_sherbrooke} and \texttt{ibm\_torino}. In Figure \ref{fig13}, each logical qubit consists of five data qubits, with four spatial syndrome bits per logical qubit. These syndrome bits, labeled as \textbf{S1}, \textbf{S2}, \textbf{S3}, and \textbf{S4}, correspond to their positions in the code. It is important to note that each \textbf{S1} from the two logical qubits corresponds locally within the perspective of each logical qubit. The same correspondence applies to the other syndrome bits (\textbf{S2}, \textbf{S3}, \textbf{S4}). The syndrome extraction circuit is performed ten times, with the corresponding syndrome having eleven syndrome rounds (\textbf{SR}). Hence, the axes represent the four blocks corresponding to syndrome nodes for a logical qubit, with each block displaying eleven syndrome rounds arranged according to their location within the code. The transversal CNOT gate is implemented in the middle of the syndrome extraction rounds ($\mathbf{SR=6}$), allowing for the observation of the correlation between the two logical qubits as depicted in the figures. In the case of \texttt{ibm\_torino} shown in Figure \ref{fig14}, the only difference is the number of data qubits per logical qubit, which is seven. 

In Figure \ref{fig13}, strong correlations are observed between the logical control and target qubits, particularly when the logical CNOT gate is introduced. Additionally, each syndrome bit from one logical qubit, including nearest neighbor syndrome bits, shows correlations with the locally corresponding syndrome bit from the other logical qubit. These correlations signify error propagation affecting the logical state and are represented differently depending on the type of errors, reflecting the propagation of Pauli operators. On the other hand, a noticeable high in error probability is observed for specific syndrome bits, such as (a) \textbf{S2} or (b) \textbf{S3}, at the logical target qubit. Although these errors can be decomposed into consecutive time errors, their actual nature may vary, including leakage or correlated errors. These differences contribute to higher overall logical error rates. These properties are also observed in Figure \ref{fig14} except for the regions where the syndrome bit \textbf{S2} is related. 

Intriguingly, there is another type of error that cause low performance of the transversal CNOT gate. Particularly in the region related to the syndrome bit \textbf{S2} within the results obtained from \texttt{ibm\_torino}, most of the pixels behave differently, specifically showing 0 values. This indicates that they have negative values according to equation \ref{eq1}. These errors in the syndrome bit \textbf{S2} were also observed across various distance structures and different rounds within the region where using the same physical qubits. These observations suggest that errors occurring during the execution of quantum circuits on the device primarily manifest locally. This localized error behavior, particularly around the syndrome bit \textbf{S2}, highlights challenges inherent to quantum processors that warrant further investigation. We hope that our demonstration will accelerate studies addressing these challenges, including understanding and mitigating inherent errors in quantum processors, which should be a key focus of ongoing research for a fault-tolerant quantum computer.

\begin{figure*}[ht]
    \centering
    \includegraphics[width=\textwidth]{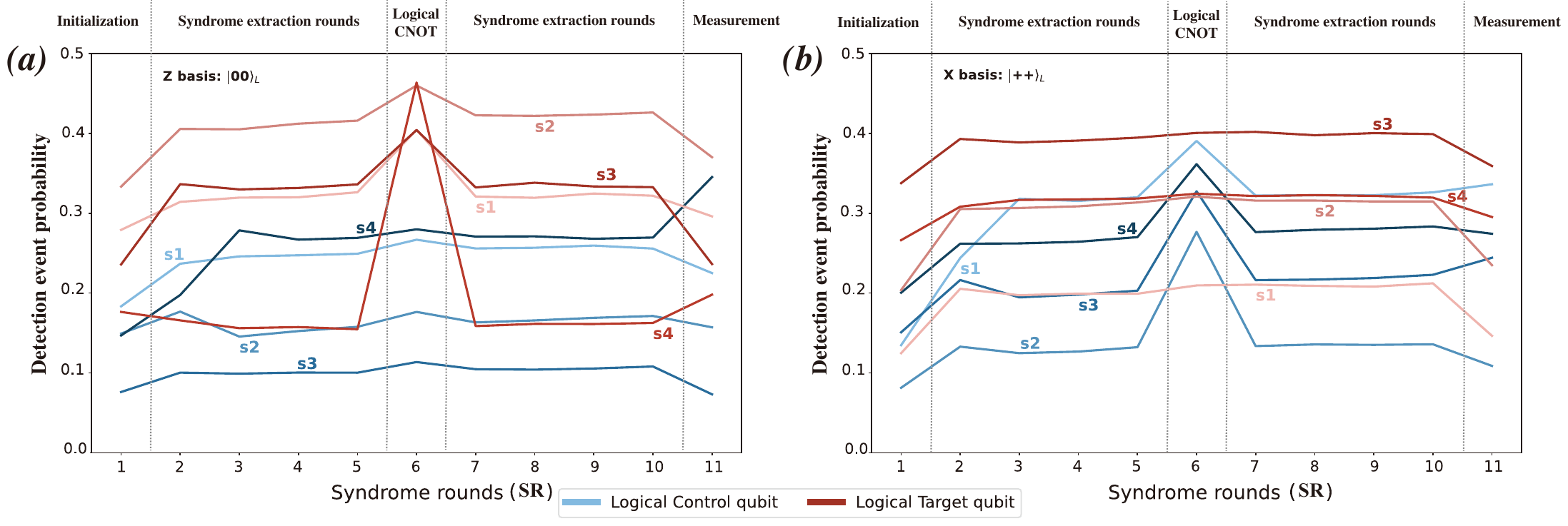}%
    \caption{ Probabilities of a detection event on each syndrome node over rounds (\texttt{ibm\_sherbrooke}). The data is obtained from the experiment with the initial state (a) $\ket{00}_L$ and (b) $\ket{++}_L$ with $\mathbf{2R=10}$ syndrome extraction rounds. The probability of the syndrome node being the detection event over time is plotted as a function of syndrome rounds. The number of syndrome rounds is eleven and steps of the quantum memory experiment can be seen based on the syndrome rounds. The transversal CNOT gate is executed at $\mathbf{SR=6}$. The red(blue) lines correspond to the probabilities of the syndrome bits of the logical control(target) qubit. The syndrome bits are labeled as \textbf{S1}, \textbf{S2}, \textbf{S3}, and \textbf{S4} according to their location in the code. The number of samples is $10^5$ to calculate the probabilities for each case. }
    \label{fig15}
\end{figure*}

\begin{figure*}[ht]
    \centering
    \includegraphics[width=\textwidth]{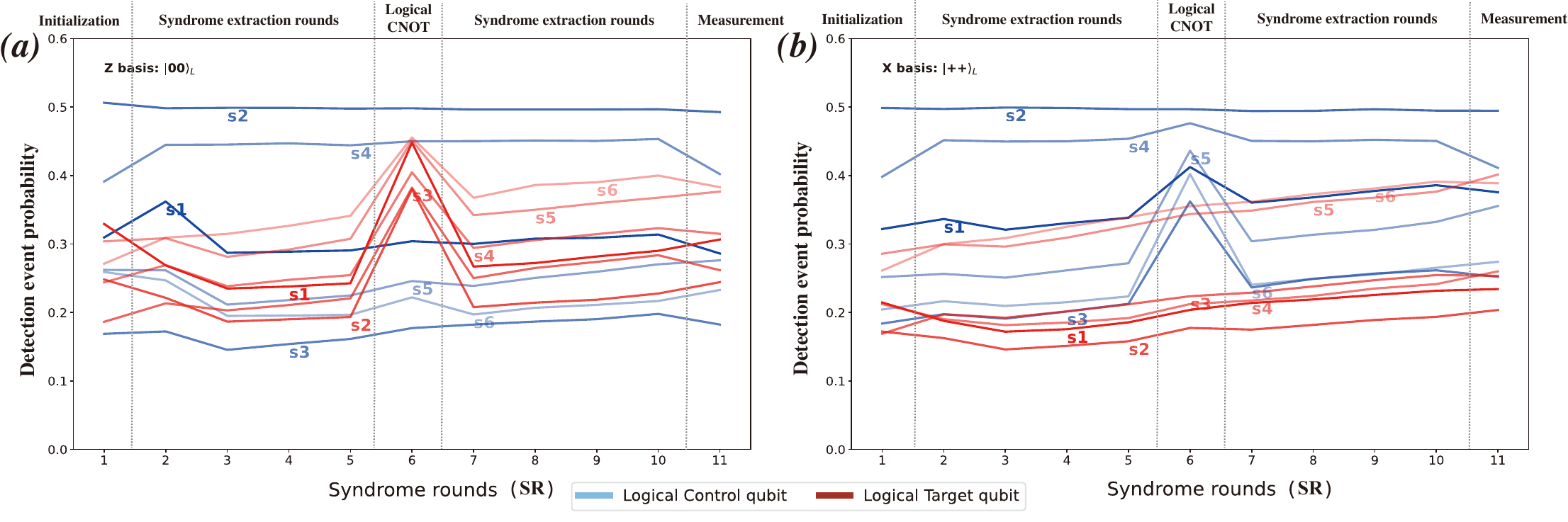}%
    \caption{ Probabilities of a detection event on each syndrome node over rounds (\texttt{ibm\_torino}). The data is obtained from the experiment with the initial state (a) $\ket{00}_L$ and (b) $\ket{++}_L$ with $\mathbf{2R=10}$ syndrome extraction rounds. The probability of the syndrome node being the detection event over time is plotted as a function of syndrome rounds. The number of syndrome rounds is eleven and steps of the quantum memory experiment can be seen based on the syndrome rounds. The transversal CNOT gate is executed at $\mathbf{SR=6}$. The red(blue) lines correspond to the probabilities of the syndrome bits of the logical control(target) qubit. The syndrome bits are labeled as \textbf{S1}, \textbf{S2}, \textbf{S3}, \textbf{S4}, \textbf{S5}, and \textbf{S6} according to their location in the code. The number of samples is $10^5$ to calculate the probabilities for each case. }
    \label{fig16}
\end{figure*}

\subsection{Detection probability}\label{app:supp5.3}

The probability of a detection event can be calculated by the ratio of how many times a syndrome bit is observed as a detection event in each syndrome round (\textbf{SR}) and the number of trials. Figure \ref{fig15} illustrates the probabilities of detection events obtained from \texttt{ibm\_sherbrooke} as a function of syndrome rounds and syndrome node labels for the $d=5$ structure on the (a) Z ($\ket{00}_L$) or (b) X ($\ket{++}_L$) basis. The blue and red lines represent the detection event probabilities for the logical control and target qubits, respectively. The logical CNOT gate is applied after five rounds of syndrome extraction circuits which is $\mathbf{SR=6}$. In the case of X errors, they propagate from the logical control qubit to the logical target qubit, whereas for Z errors, the propagation occurs in the opposite direction. Notably, the detection event probabilities for all syndrome bits on the logical target (control) qubit increase significantly when the logical qubits are in the Z (X) basis at $\mathbf{SR=6}$. Furthermore, the probabilities for syndrome bits on the other logical qubit also temporally increase when the logical CNOT gate is implemented. This increase is attributed to the use of auxiliary qubits in implementing the transversal CNOT gate, which introduces the possibility of errors associated with these auxiliary qubits during the application of the logical CNOT gate. Consequently, the detection event probabilities for syndrome nodes on both logical qubits with Z (X) stabilizers increase, reflecting the impact of Gate-Flow errors. It is also similarly observed from \texttt{ibm\_torino} except for the \textbf{S2} syndrome bit with inherent errors as shown in Figure \ref{fig16}.

\subsection{Logical error rates}\label{app:supp5.4}

\begin{figure*}[ht]
    \centering
    \includegraphics[width=\textwidth]{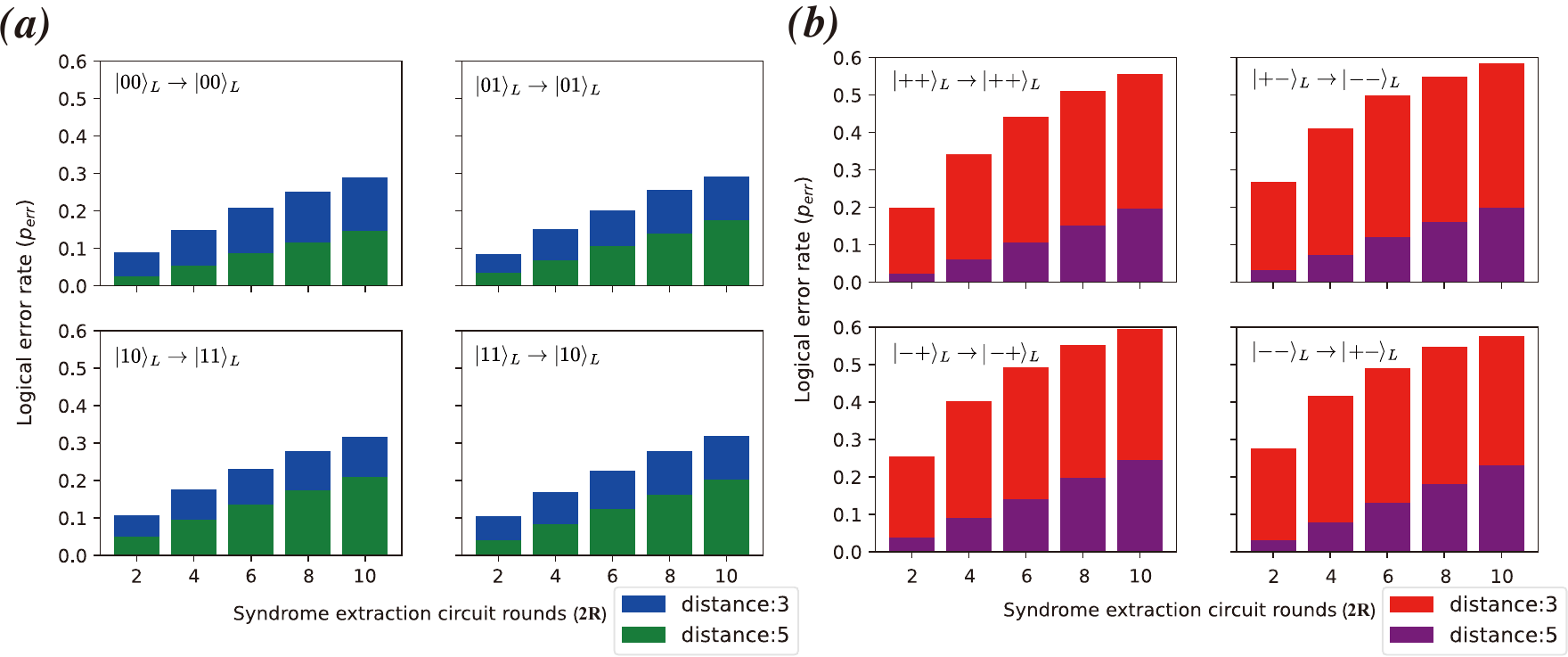}
    \caption{ Bar charts showing the logical error rates (\texttt{ibm\_sherbrooke}). The logical qubits initialized in the (a) Z or (b) X basis, calculated based on data sampled $10^5$ times for each case using the quantum memory experiment. The logical error rates are displayed as a function of distances and the number of rounds of the syndrome extraction circuit. The logical errors for distances 3, and 5 are plotted non-cumulatively across the syndrome rounds. The state of the logical qubits can be prepared as one of four states for each basis. Subfigures illustrate the case for a certain initial state. }
    \label{fig17}
\end{figure*}

Experiments are conducted with rounds of the corresponding syndrome extraction circuit, initializing the logical qubits in four different states for each basis. Here, $|\psi_C \psi_T \rangle_L = |\psi_C\rangle_L \otimes|\psi_T\rangle_L$ denotes the product state of logical control ($\ket{\psi_C}_L$) and target ($\ket{\psi_T}_L$) qubits. For example, when using Z stabilizers, the computational states of two logical qubits can be prepared as $\{\ket{00}_L, \ket{01}_L, \ket{10}_L, \ket{11}_L\}$. Applying the logical CNOT gate to these states transforms them into $\{\ket{00}_L, \ket{01}_L, \ket{11}_L, \ket{10}_L\}$, respectively. Similarly, for the X basis, the initial states are prepared as one of $\{\ket{++}_L, \ket{+-}_L, \ket{-+}_L, \ket{--}_L\}$, which change into $\{\ket{++}_L, \ket{--}_L, \ket{-+}_L, \ket{+-}_L\}$, respectively. 

Logical error rates ($p_{err}$) are calculated as infidelity ($1-F_{\psi_C,\psi_T}$), where $F_{\psi_C,\psi_T}$ is the fidelity when the output state is $\ket{\psi_C \psi_T}_L$ after the logical CNOT gate has been executed. After decoding errors and correcting the two logical qubit states, the fidelity of the corrected logical qubits is computed using the logical qubits' Pauli operators \cite{trapped_ion_transversal}. For the Z basis, the fidelity calculation is as follows:

\begin{align}
F_{00}&=\frac{1}{4}(1+\langle Z_L^C \rangle+\langle Z_L^T \rangle+\langle Z_L^C Z_L^T \rangle ) \\
F_{01}&=\frac{1}{4}(1+\langle Z_L^C \rangle-\langle Z_L^T \rangle-\langle Z_L^C Z_L^T \rangle ) \\
F_{10}&=\frac{1}{4}(1-\langle Z_L^C \rangle+\langle Z_L^T \rangle-\langle Z_L^C Z_L^T \rangle ) \\
F_{11}&=\frac{1}{4}(1-\langle Z_L^C \rangle-\langle Z_L^T \rangle+\langle Z_L^C Z_L^T \rangle )
\label{eq2}
\end{align}

The fidelity calculations involve measuring the expectation values of logical Pauli operators. Specifically, $\langle Z_L^C \rangle$, $\langle Z_L^T \rangle$, and $\langle Z_L^C Z_L^T \rangle$ represent the expectation values of the logical Z operator on the logical control qubit, the logical target qubit, and both logical qubits, respectively. The same process can be applied to the logical qubits in the X basis when dealing with the phase-flip code with flag qubits.

\begin{align}
F_{++}&=\frac{1}{4}(1+\langle X_L^C \rangle+\langle X_L^T \rangle+\langle X_L^C X_L^T \rangle ) \\
F_{+-}&=\frac{1}{4}(1+\langle X_L^C \rangle-\langle X_L^T \rangle-\langle X_L^C X_L^T \rangle ) \\
F_{-+}&=\frac{1}{4}(1-\langle X_L^C \rangle+\langle X_L^T \rangle-\langle X_L^C X_L^T \rangle ) \\
F_{--}&=\frac{1}{4}(1-\langle X_L^C \rangle-\langle X_L^T \rangle+\langle X_L^C X_L^T \rangle )
\label{eq3}
\end{align}

\begin{figure*}[ht]
    \centering
    \includegraphics[width=\textwidth]{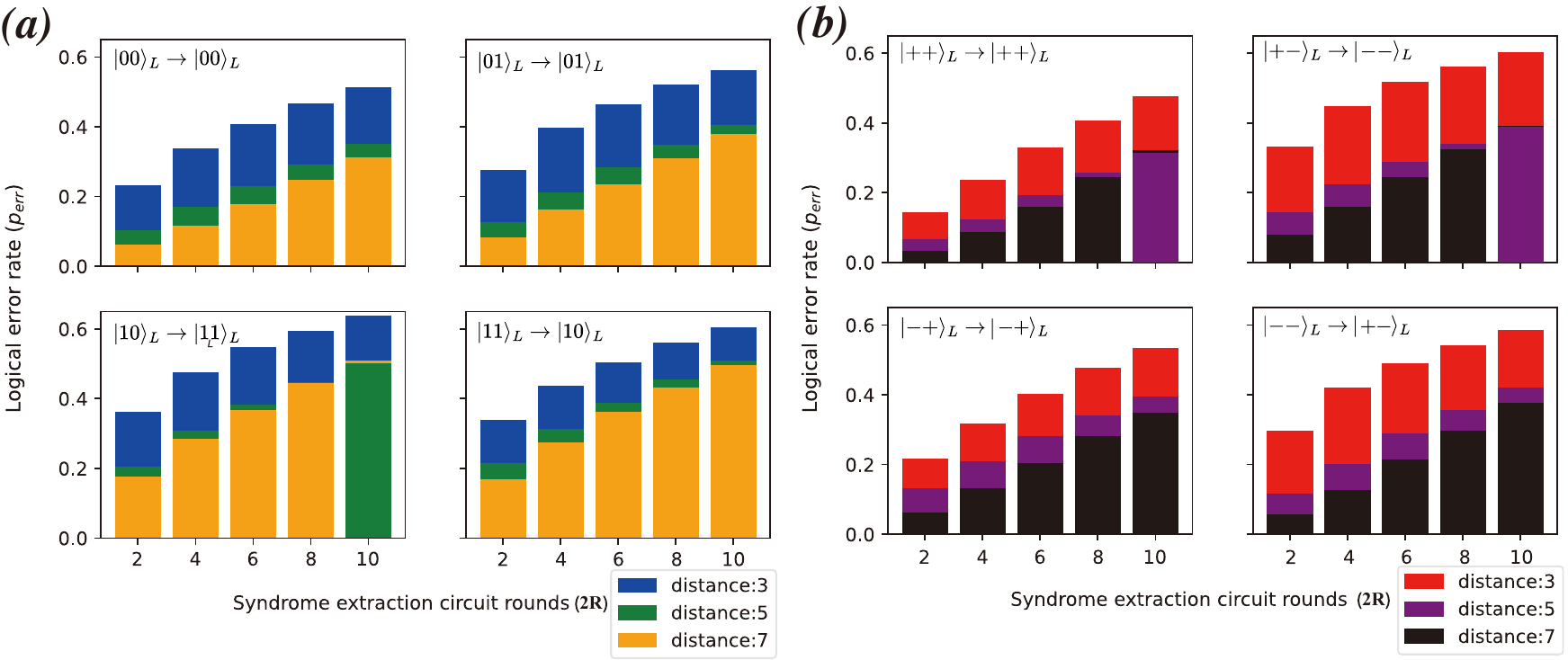}
    \caption{ Bar charts showing the logical error rates (\texttt{ibm\_torino}). The logical qubits initialized in the (a) Z or (b) X basis, calculated based on data sampled $10^5$ times for each case using the quantum memory experiment. The logical error rates are displayed as a function of distances and the number of rounds of the syndrome extraction circuit. The logical errors for distances 3, 5, and 7 are plotted non-cumulatively across the syndrome rounds. The state of the logical qubits can be prepared as one of four states for each basis. Subfigures illustrate the case for a certain initial state. }
    \label{fig18}
\end{figure*}

Figure \ref{fig17} displays the logical error rates for four states with non-cumulatively bar charts across the syndrome rounds, obtained from \texttt{ibm\_sherbrooke}. The reduction in logical errors regardless of the initial state and basis is observed when more physical qubits are utilized. However, we observe that logical errors for the X basis are higher than those for the Z basis across all syndrome extraction rounds. However, the $d=5$ structure exhibits a lower logical error rate than the $d=3$ structure for all cases. The same figure obtained from \texttt{ibm\_torino} is shown in Figure \ref{fig18}. Even though, in some cases, the larger structure does not have lower logical errors than the smaller one the trends of logical error rates across all syndrome extraction circuit rounds indicate the error suppression overall. This could be attributed to the inherent errors in the \textbf{S2} syndrome bit region which has covered in Appendix \ref{app:supp5.3}.

\begin{figure*}[ht]
    \centering
    \includegraphics[width=\textwidth]{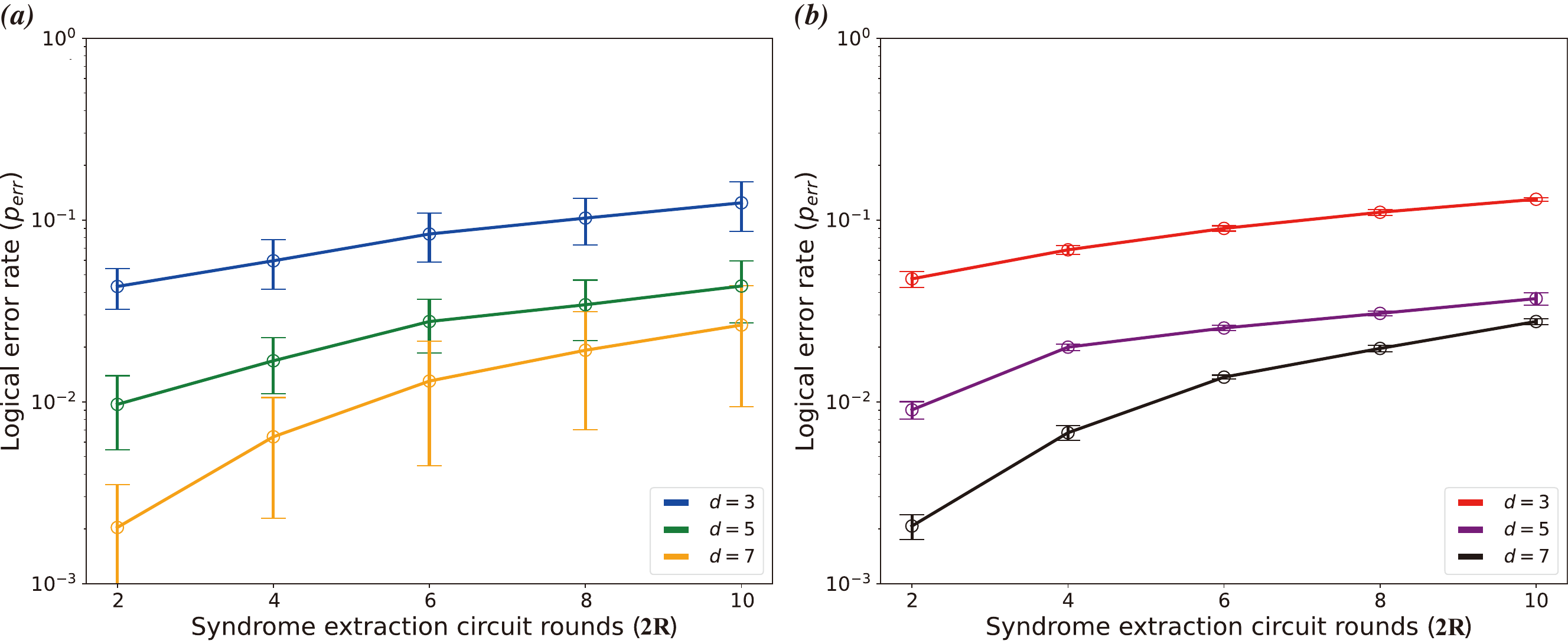}
    \caption{ Logical error rates (\texttt{Fakewashington}). The code considers (a) X errors with Z stabilizers or (b) Z errors with X stabilizers. To assess the logical error rates, the MWPM decoder is used. The blue (red), green (purple), and orange (black) lines correspond to structure sizes of 21, 39, and 57 physical qubits correcting X (Z) errors, respectively. The lines are obtained by sampling the quantum memory experiment $2 \times 10^4$ times on the Fakewashington simulator. The average logical error rates for four basis states are depicted with their standard deviation as a function of the number of syndrome extraction circuit rounds, and structure size. As the number of physical qubits increases, the logical error rates decrease for both X and Z error cases. }
    \label{fig19}
\end{figure*}

Figure \ref{fig19} shows the logical error rates as a function of the number of syndrome extraction circuits and distance, obtained from the data from the classical simulator (\texttt{Fakewashington}). The structure with more physical qubits has a lower logical error rate for all syndrome circuit rounds. As we increase the number of syndrome extraction circuit rounds, the gap between the logical error rates with different distances narrows for both bases. This behavior is also observed in the results from \texttt{ibm\_torino}. This is attribute to that the increase in the possibility of Gate-Flow errors, which may lead to ambiguity for a decoder when calculating a correction operator due to correlated errors, especially for the MWPM decoder, is associated with the number of syndrome extraction cycles per the logical CNOT gate.

\begin{figure*}[ht]
    \centering
    \includegraphics[width=\textwidth]{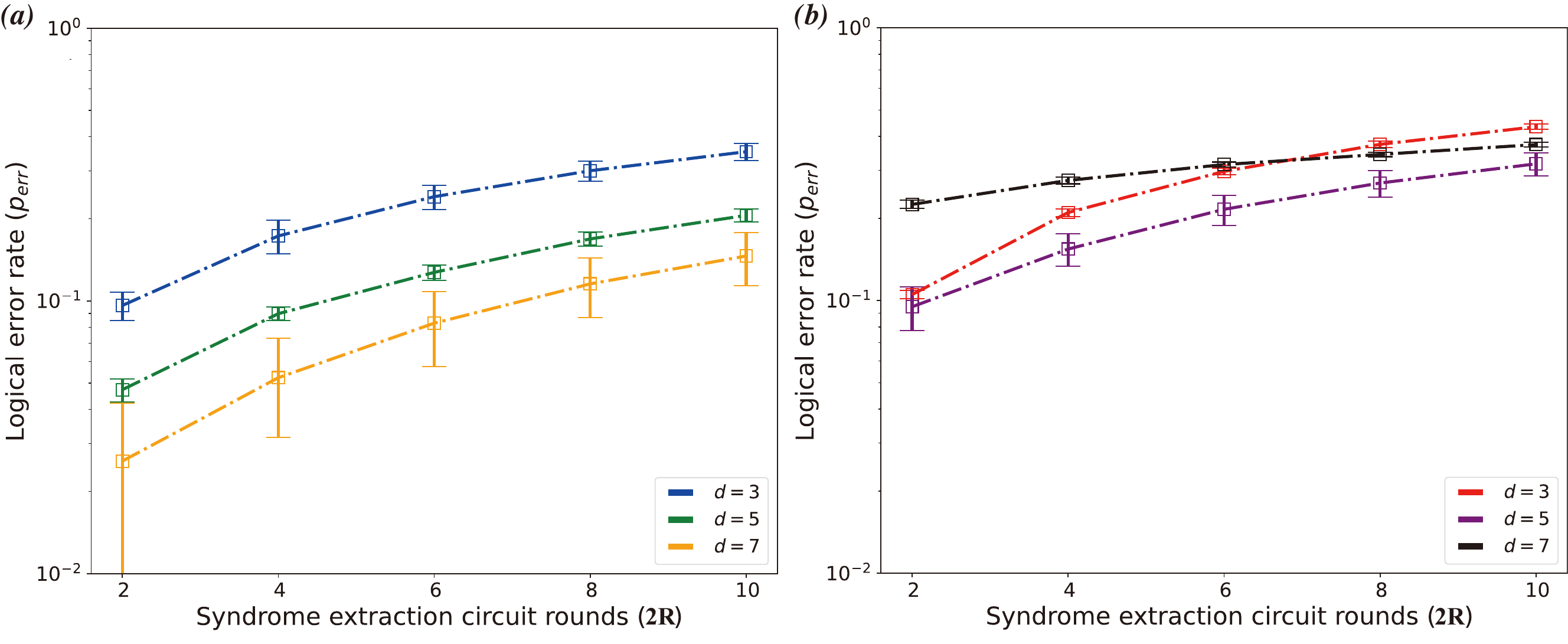}
    \caption{ Logical error rates (\texttt{brisbane}). The code considers (a) X errors with Z stabilizers or (b) Z errors with X stabilizers. To assess the logical error rates, the MWPM decoder is used. The blue (red), green (purple), and orange (black) lines correspond to structure sizes of 21, 39, and 57 physical qubits correcting X (Z) errors, respectively. The lines are obtained by sampling the quantum memory experiment $10^5$ times on \texttt{brisbane}. The average logical error rates for four basis states are depicted with their standard deviation as a function of the number of syndrome extraction circuit rounds, and structure size. As the number of physical qubits increases, the logical error rates decrease for both X and Z error cases. }
    \label{fig20}
\end{figure*}

Figure \ref{fig20} presents the same graph as Figure \ref{fig19}, but obtained from \texttt{ibm\_brisbane}. When the logical qubits are prepared in the Z basis, the logical error rates notably decrease as we increase the number of physical qubits. However, while there is a decrease in logical error rates between $d=3$ and $d=5$, the $d=7$ structure exhibits higher logical error rates across the syndrome extraction rounds than that of the $d=5$ structure. It could be related to the biased errors during the execution of the logical CNOT gate as discussed in Appendix \ref{app:supp5.2}.

\clearpage
\section{Further figures}\label{app:supp6}
\begin{table*}[htb]
  \resizebox{\textwidth}{!}{
    \begin{tabular}{||c | c | c | c | c | c||} 
     \hline
     Qubit & T1($\mu$s) & T2($\mu$s) & Readout Error Rate& SQ Gate Error Rate& TQ Gate Error Rate\\ 
     \hline\hline
    0 & 248.48 & 368.92 & 0.0133 & 0.00017 & [0.0137, 0.0048] \\ 
     \hline 
    1 & 264.36 & 317.35 & 0.0132 & 0.00014 & [0.0048, 0.0041] \\ 
     \hline 
    2 & 284.58 & 306.75 & 0.008 & 0.00038 & [0.0087, 0.0041] \\ 
     \hline 
    3 & 216.28 & 67.9 & 0.0065 & 0.00019 & [0.0051, 0.0087] \\ 
     \hline 
    4 & 430.15 & 78.68 & 0.008 & 0.00015 & [0.0051, 0.0051, 0.0053] \\ 
     \hline 
    5 & 330.82 & 290.9 & 0.0137 & 0.00028 & [0.0067, 0.0053] \\ 
     \hline 
    6 & 318.33 & 183.28 & 0.0136 & 0.00018 & [0.0086, 0.0067, 0.0051] \\ 
     \hline 
    7 & 271.58 & 287.72 & 0.12 & 0.00018 & [0.005, 0.0051] \\ 
     \hline 
    8 & 308.95 & 262.67 & 0.0134 & 0.00011 & [0.005, 0.0032] \\ 
     \hline 
    9 & 226.93 & 41.98 & 0.0148 & 0.00049 & [0.0137, 0.0136] \\ 
     \hline 
    10 & 493.14 & 381.62 & 0.0202 & 0.00013 & [0.0051, 0.0085] \\ 
     \hline 
    11 & 373.5 & 177.92 & 0.0068 & 0.00019 & [0.0074, 0.0032] \\ 
     \hline 
    12 & 227.92 & 60.3 & 0.0083 & 0.00038 & [0.0079, 0.0136, 0.0082] \\ 
     \hline 
    13 & 363.33 & 81.61 & 0.0131 & 0.00018 & [0.0051, 0.0082] \\ 
     \hline 
    14 & 254.07 & 186.99 & 0.0081 & 0.00031 & [0.0095, 0.0051, 0.0062] \\ 
     \hline 
    15 & 238.31 & 156.11 & 0.0166 & 0.00034 & [0.0095, 0.0076] \\ 
     \hline 
    16 & 139.01 & 91.39 & 0.0278 & 0.00028 & [0.0076, 0.0155, 0.0085] \\ 
     \hline 
    17 & 307.58 & 261.39 & 0.0529 & 0.00055 & [0.0155, 0.0055] \\ 
     \hline 
    18 & 249.95 & 245.07 & 0.0066 & 0.00015 & [0.0045, 0.0074, 0.0055] \\ 
     \hline 
    19 & 274.84 & 16.84 & 0.0172 & 0.00027 & [0.0045, 0.0056] \\ 
     \hline 
    20 & 174.69 & 202.76 & 0.0082 & 0.00019 & [0.0056, 0.0074, 0.011] \\ 
     \hline 
    \end{tabular}
  }\caption{ Calibration table values for 21 physical qubits used for experiments (\texttt{ibm\_brisbane}). The values obtained from the calibration table correspond to the experiment for $|00\rangle_L$ with ten rounds. The table lists the properties of physical qubits for the $d=3$ structure selected within the hardware. These include T1, and T2 time, as well as error rates of readout, single- (SQ) and two-qubit (TQ) gates. More than one two-qubit gate can be implemented and their error rates are listed accordingly. }
  \label{table_brisbane}
\end{table*}

\begin{figure*}[t]
    \centering
    \includegraphics[width=\textwidth]{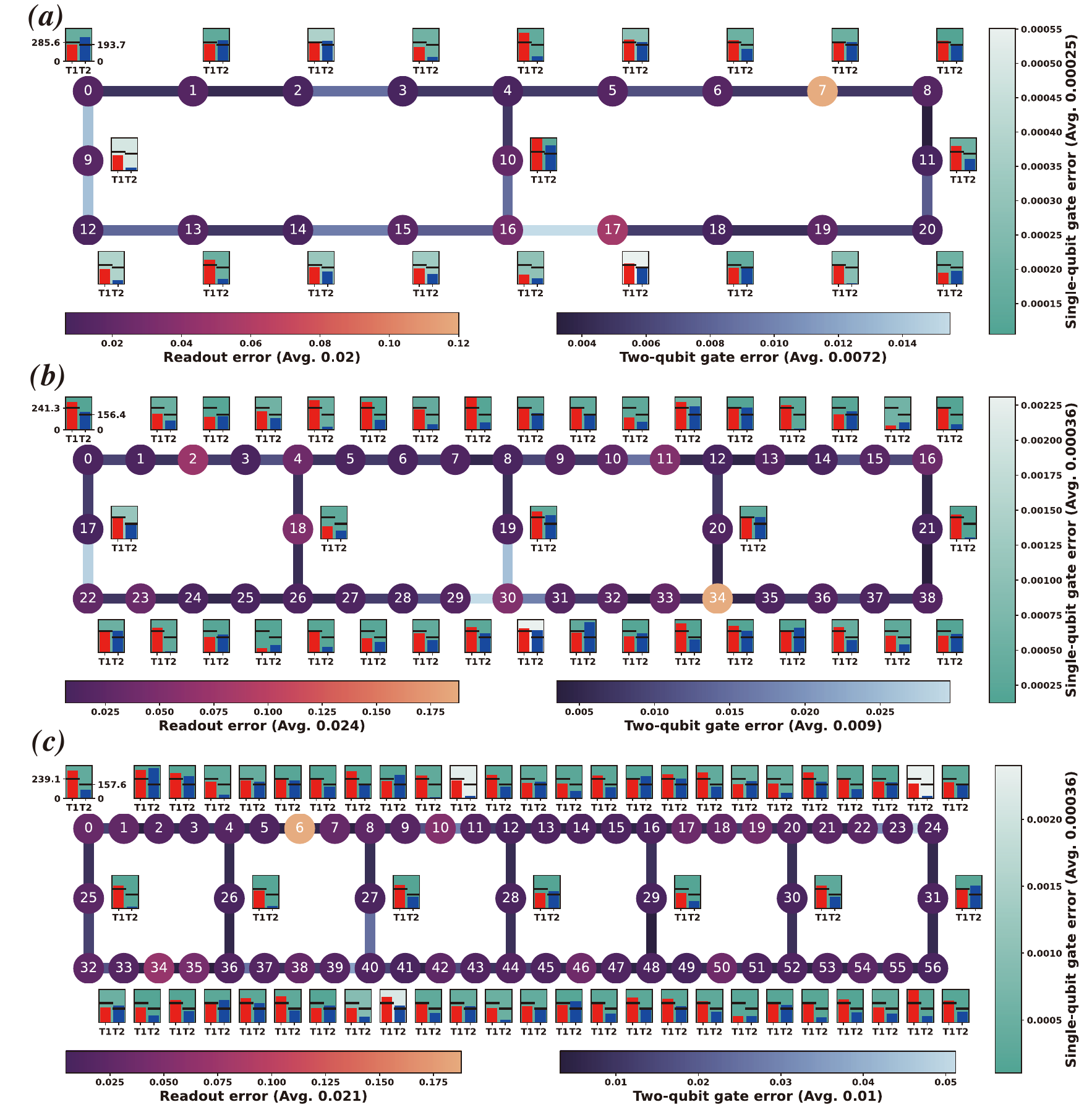}
    \caption{ Hardware specifications of physical qubits with colormap and bar charts (\texttt{ibm\_brisbane}). The graph with subplots shows the physical qubits' specification of the structure in (a) $d=3$, (b) $d=5$, and (c) $d=7$ with $\ket{00}_L$ state. Each node and edge correspond to the physical qubit and connectivity of a two-qubit gate. The error rates of readout and two-qubit gate (ECR) are displayed with colors. The subplot next to every node shows T1 and T2 times corresponding to their physical qubit. The average time of T1 and T2 are plotted as the black lines on the left and right sides. Their values can be seen on the subplot of the 0th physical qubit. The background of a subplot is colored corresponding to the error rate of the single-qubit gate. }
    \label{fig21}
\end{figure*}

\begin{table*}[htb]
  \resizebox{\textwidth}{!}{
    \begin{tabular}{||c | c | c | c | c | c||} 
     \hline
     Qubit & T1($\mu$s) & T2($\mu$s) & Readout Error Rate& SQ Gate Error Rate& TQ Gate Error Rate\\ 
     \hline\hline
    0 & 142.71 & 116.51 & 0.085 & 0.00053 & [0.0044, 0.0083] \\ 
     \hline 
    1 & 132.53 & 35.6 & 0.0143 & 0.00065 & [0.0082, 0.0044] \\ 
     \hline 
    2 & 127.68 & 203.37 & 0.0382 & 0.00037 & [0.0082, 0.0164] \\ 
     \hline 
    3 & 204.95 & 136.37 & 0.0077 & 0.00021 & [0.0164, 0.0044] \\ 
     \hline 
    4 & 135.37 & 25.33 & 0.0393 & 0.00067 & [0.0063, 0.0079, 0.0044] \\ 
     \hline 
    5 & 242.65 & 76.18 & 0.015 & 0.00021 & [0.0063, 0.0044] \\ 
     \hline 
    6 & 124.4 & 187.37 & 0.02167 & 0.00017 & [0.0062, 0.0029, 0.0044] \\ 
     \hline 
    7 & 38.45 & 45.24 & 0.0321 & 0.00213 & [0.0062, 0.0071] \\ 
     \hline 
    8 & 205.84 & 157.08 & 0.0315 & 0.00024 & [0.0049, 0.0071, 0.0023] \\ 
     \hline 
    9 & 232.22 & 128.96 & 0.0126 & 0.00022 & [0.0034, 0.0083] \\ 
     \hline 
    10 & 32.42 & 45.8 & 0.0131 & 0.00087 & [0.0085, 0.0079] \\ 
     \hline 
    11 & 248.0 & 160.5 & 0.0117 & 0.0002 & [0.0046, 0.0023] \\ 
     \hline 
    12 & 215.6 & 122.98 & 0.0282 & 0.00033 & [0.0053, 0.0034] \\ 
     \hline 
    13 & 263.96 & 194.63 & 0.0172 & 0.00038 & [0.0053, 0.0176] \\ 
     \hline 
    14 & 160.04 & 214.99 & 0.0129 & 0.00016 & [0.0024, 0.0176, 0.0043] \\ 
     \hline 
    15 & 174.22 & 200.89 & 0.0102 & 0.00018 & [0.0042, 0.0024] \\ 
     \hline 
    16 & 164.92 & 150.59 & 0.017 & 0.00032 & [0.0042, 0.0085, 0.0085] \\ 
     \hline 
    17 & 23.84 & 22.13 & 0.0243 & 0.00117 & [0.0085, 0.0068] \\ 
     \hline 
    18 & 219.47 & 64.31 & 0.0163 & 0.00027 & [0.0033, 0.0029, 0.0068] \\ 
     \hline 
    19 & 265.79 & 111.69 & 0.0119 & 0.00032 & [0.0045, 0.0033] \\ 
     \hline 
    20 & 158.13 & 187.47 & 0.0143 & 0.00019 & [0.0045, 0.0086, 0.0046] \\ 
     \hline 
    \end{tabular}
  }\caption{ Calibration table values for 21 physical qubits used for experiments (\texttt{ibm\_torino}). The values obtained from the calibration table correspond to the experiment for $|00\rangle_L$ with ten rounds. The table lists the properties of physical qubits for the $d=3$ structure selected within the hardware. These include T1, and T2 time, as well as error rates of readout, single- (SQ) and two-qubit (TQ) gates. More than one two-qubit gate can be implemented and their error rates are listed accordingly. }
  \label{table_torino}
\end{table*}

\begin{figure*}[t]
    \centering
    \includegraphics[width=\textwidth]{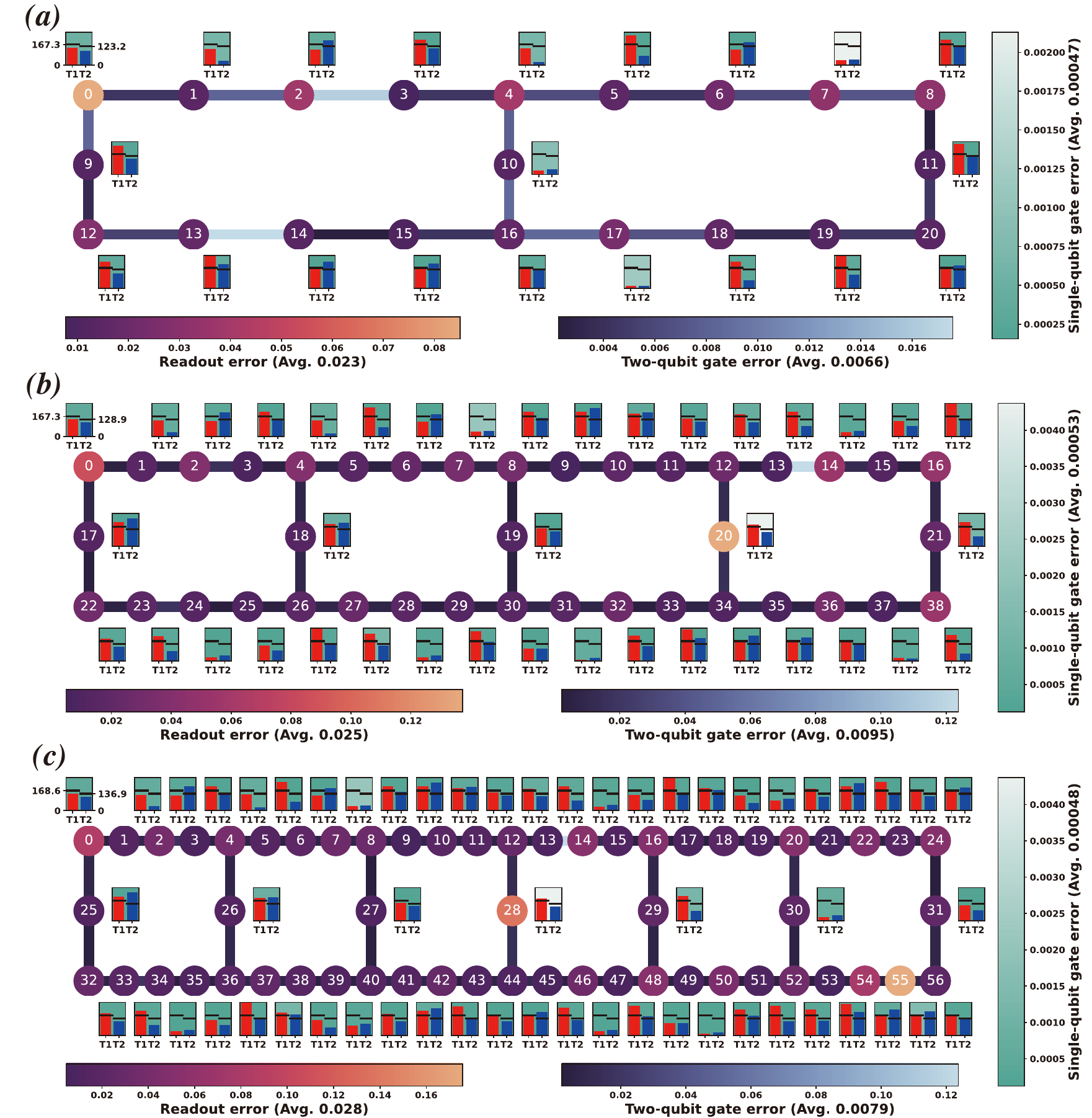}
    \caption{ Hardware specifications of physical qubits with colormap and bar charts (\texttt{ibm\_torino}). The graph with subplots shows the physical qubits' specification of the structure in (a) $d=3$, (b) $d=5$, and (c) $d=7$ with $\ket{00}_L$ state. Each node and edge correspond to the physical qubit and connectivity of a two-qubit gate (CZ). The error rates of readout and two-qubit gate are displayed with colors. The subplot next to every node shows T1 and T2 times corresponding to their physical qubit. The average time of T1 and T2 are plotted as the black lines on the left and right sides. Their values can be seen on the subplot of the 0th physical qubit. The background of a subplot is colored corresponding to the error rate of the single-qubit gate. }
    \label{fig22}
\end{figure*}

\begin{table*}[htb]
  \resizebox{\textwidth}{!}{
    \begin{tabular}{||c | c | c | c | c | c||} 
     \hline
     Qubit & T1($\mu$s) & T2($\mu$s) & Readout Error Rate& SQ Gate Error Rate& TQ Gate Error Rate\\ 
     \hline\hline
    0 & 64.12 & 66.78 & 0.0124 & 0.00036 & [0.0123, 0.0087, 0.017] \\ 
     \hline 
    1 & 142.83 & 145.96 & 0.0042 & 0.00021 & [0.0123, 0.0085] \\ 
     \hline 
    2 & 126.28 & 155.98 & 0.0054 & 0.00018 & [0.0165, 0.0076, 0.0085] \\ 
     \hline 
    3 & 139.68 & 196.76 & 0.0297 & 0.00019 & [0.0153, 0.0076] \\ 
     \hline 
    4 & 98.66 & 54.68 & 0.0554 & 0.00033 & [0.0153, 0.0097, 0.0179] \\ 
     \hline 
    5 & 99.17 & 82.35 & 0.0067 & 0.00026 & [0.0097, 0.0097] \\ 
     \hline 
    6 & 109.6 & 69.53 & 0.0196 & 0.00019 & [0.0416, 0.0097, 0.0062] \\ 
     \hline 
    7 & 108.44 & 44.67 & 0.0153 & 0.00021 & [0.0062, 0.0096] \\ 
     \hline 
    8 & 60.6 & 98.18 & 0.007 & 0.00028 & [0.0096, 0.0103] \\ 
     \hline 
    9 & 105.36 & 118.16 & 0.0124 & 0.00023 & [0.0087, 0.0364] \\ 
     \hline 
    10 & 69.12 & 15.95 & 0.0456 & 0.00133 & [0.0232, 0.0179] \\ 
     \hline 
    11 & 74.22 & 63.46 & 0.0055 & 0.00027 & [0.0286, 0.0103] \\ 
     \hline 
    12 & 140.53 & 122.58 & 0.0353 & 0.00034 & [0.0247, 0.0398, 0.0364] \\ 
     \hline 
    13 & 87.4 & 127.4 & 0.0267 & 0.00066 & [0.0398, 0.013] \\ 
     \hline 
    14 & 113.57 & 178.72 & 0.0206 & 0.00218 & [0.0204, 0.013, 0.0109] \\ 
     \hline 
    15 & 160.15 & 284.58 & 0.0063 & 0.00024 & [0.009, 0.0109] \\ 
     \hline 
    16 & 98.16 & 46.0 & 0.0286 & 0.0003 & [0.0232, 0.0076, 0.009] \\ 
     \hline 
    17 & 88.71 & 18.09 & 0.0096 & 0.00031 & [0.0076, 0.0109] \\ 
     \hline 
    18 & 106.8 & 128.09 & 0.0157 & 0.00024 & [0.0111, 0.0075, 0.0109] \\ 
     \hline 
    19 & 84.11 & 91.22 & 0.0099 & 0.00026 & [0.0075, 0.0132] \\ 
     \hline 
    20 & 57.8 & 62.75 & 0.0218 & 0.02651 & [0.0132, 0.0286] \\ 
     \hline 
    \end{tabular}
  }\caption{ Calibration table values for 21 physical qubits used for experiments (\texttt{Fakewashington}). The values obtained from the calibration table correspond to the experiment for $|00\rangle_L$ with ten rounds. The table lists the properties of physical qubits for the $d=3$ structure selected within the hardware. These include T1, and T2 time, as well as error rates of readout, single- (SQ) and two-qubit (TQ) gates. More than one two-qubit gate can be implemented and their error rates are listed accordingly. }
  \label{table_fakewashington}
\end{table*}

\begin{figure*}[t]
    \centering
    \includegraphics[width=\textwidth]{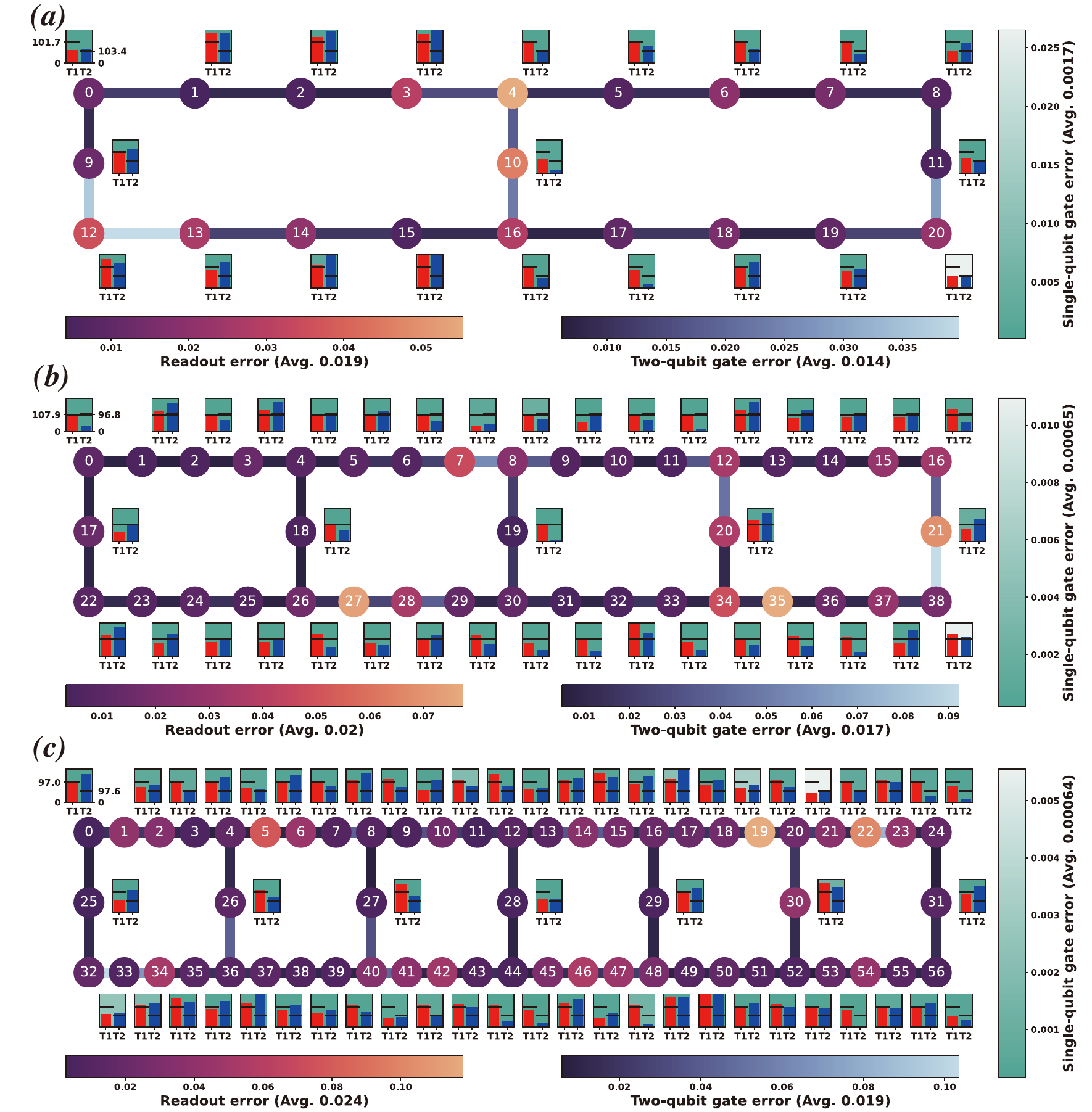}
    \caption{ Classical simulator specifications of physical qubits with colormap and bar charts (\texttt{Fakewashington}). The graph with subplots shows the physical qubits' specification of the structure in (a) $d=3$, (b) $d=5$, and (c) $d=7$ with $\ket{00}_L$ state. Each node and edge correspond to the physical qubit and connectivity of a two-qubit gate (CX). The error rates of readout and two-qubit gate are displayed with colors. The subplot next to every node shows T1 and T2 times corresponding to their physical qubit. The average time of T1 and T2 are plotted as the black lines on the left and right sides. Their values can be seen on the subplot of the 0th physical qubit. The background of a subplot is colored corresponding to the error rate of the single-qubit gate. }
    \label{fig23}
\end{figure*}

\begin{figure*}[t]
    \centering
    \includegraphics[width=\textwidth]{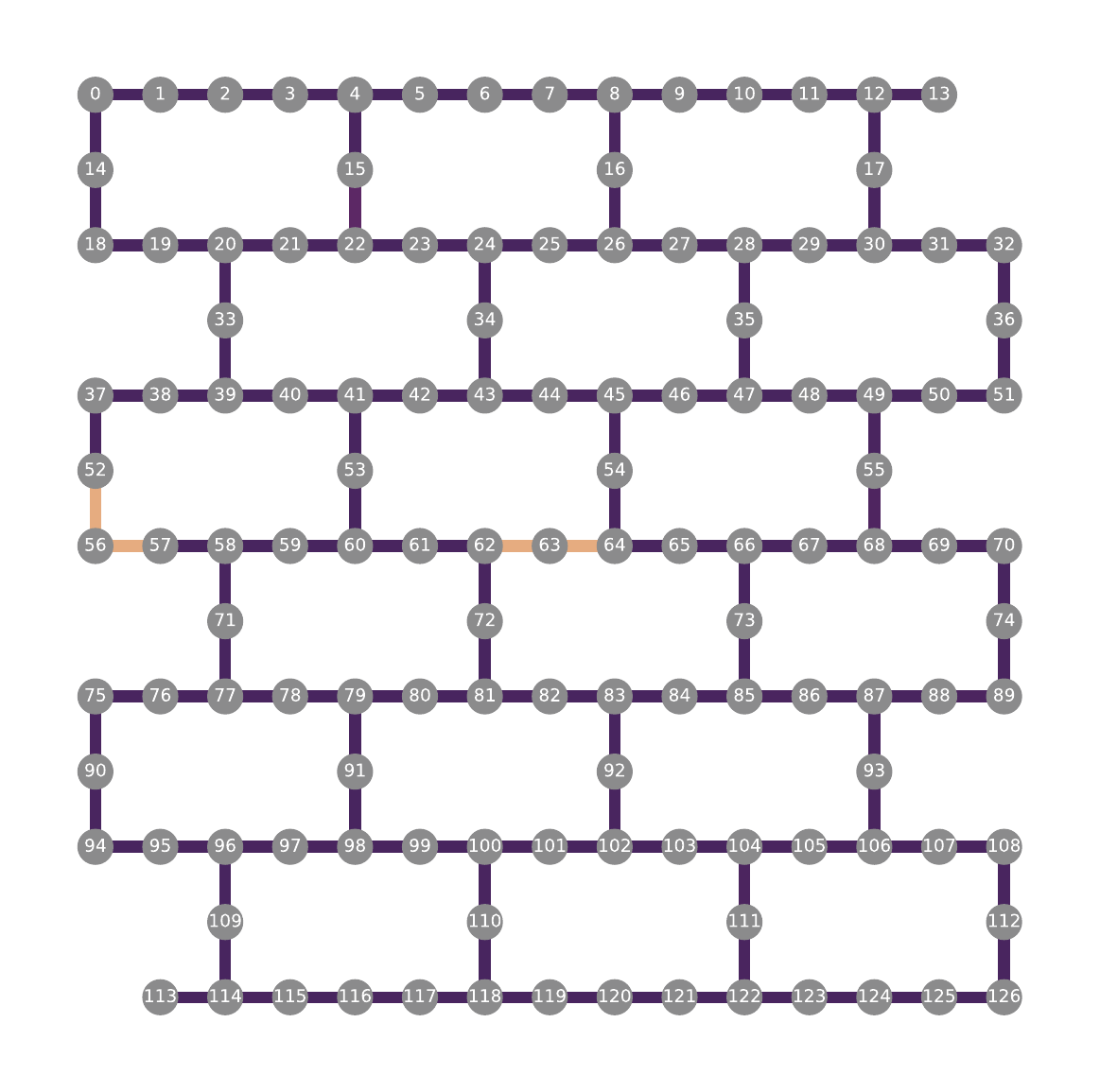}
    \caption{ Connectivity of \texttt{ibm\_sherbrooke}. Each node and edge correspond to a qubit and its connectivity with other nearest neighbor qubits, respectively. The edges highlighted in orange represent the connections where the two-qubit gate (ECR) has an error rate of 1. In contrast, other edges, indicated in purple, have error rates lower than 1. }
    \label{fig24}
\end{figure*}


\begin{thebibliography}{99}

\bibitem{NISQ}
J, Preskill.
\newblock Quantum Computing in the NISQ era and beyond.
\newblock Quantum 2, 79, \href{http://dx.doi.org/10.22331/q-2018-08-06-79}{http://dx.doi.org/10.22331/q-2018-08-06-79} (2018).

\bibitem{heavy-hexagon}
Kim, Y. et al.
\newblock  Evidence for the utility of quantum computing before fault tolerance. 
\newblock Nature 618, 500–505, \href{https://doi.org/10.1038/s41586-023-06096-3}{https://doi.org/10.1038/s41586-023-06096-3} (2023).

\bibitem{sycamore}
Arute, Frank, et al. 
\newblock Quantum supremacy using a programmable superconducting processor. 
\newblock Nature 574.7779, 505-510, \href{https://doi.org/10.1038/s41586-019-1666-5}{https://doi.org/10.1038/s41586-019-1666-5} (2019).

\bibitem{quantinuum}
Moses, S. A, et al.
\newblock  A Race-Track Trapped-Ion Quantum Processor.
\newblock Phys. Rev. {\bf X13} {041052-041077}, \href{https://doi.org/10.1103/PhysRevX.13.041052}{https://doi.org/10.1103/PhysRevX.13.041052} (2023).

\bibitem{quantinuum2}
M. DeCross, et al. 
\newblock The computational power of random quantum circuits in arbitrary geometries. 
\newblock Preprint at  \href{https://arxiv.org/abs/2406.02501}{https://arxiv.org/abs/2406.02501} (2024).

\bibitem{neutral}
Wurtz, Jonathan, et al. 
\newblock Aquila: QuEra's 256-qubit neutral-atom quantum computer.
\newblock Preprint at \href{https://arxiv.org/abs/2306.11727}{https://arxiv.org/abs/2306.11727} (2023).

\bibitem{rsa}
Gidney, C., and Ekerå, M.
\newblock How to factor 2048 bit RSA integers in 8 hours using 20 million noisy qubits. 
\newblock Quantum, 5, 433, \href{https://doi.org/10.22331/q-2021-04-15-433}{https://doi.org/10.22331/q-2021-04-15-433} (2021).

\bibitem{gottesman}
Gottesman, D.
\newblock Stabilizer Codes and Quantum Error Correction.
\newblock PhD thesis, California Institute of Technology, preprint at \href{https://arxiv.org/abs/quant-ph/9705052}{https://arxiv.org/abs/quant-ph/9705052} (1997).

\bibitem{surface}
Fowler, A. G., Mariantoni, M., Martinis, J. M., Cleland, A. N. 
\newblock Surface codes: towards practical large-scale quantum computation. 
\newblock Phys. Rev. A 86, 032324, \href{https://doi.org/10.1103/PhysRevA.86.032324}{https://doi.org/10.1103/PhysRevA.86.032324} (2012).

\bibitem{qecmemory}
Terhal, B. M. 
\newblock Quantum error correction for quantum memories. 
\newblock Rev. Mod. Phys. 87, 307, \href{https://doi.org/10.1103/RevModPhys.87.307}{https://doi.org/10.1103/RevModPhys.87.307} (2015).

\bibitem{threshold} 
E. Knill, R. Laflamme, and W. H. Zurek
\newblock Resilient quantum computation: error models and thresholds
\newblock Proceedings of the Royal Society of London. Series A: Mathematical, Physical and Engineering Sciences 454, 365, \href{https://doi.org/10.1098/rspa.1998.0166}{https://doi.org/10.1098/rspa.1998.0166} (1998).

\bibitem{threshold2} 
J. Preskill.
\newblock Reliable quantum computers.
\newblock Proceedings of the Royal Society of London. Series A: Mathematical, Physical and Engineering Sciences 454.1969:385-410, \href{https://doi.org/10.1098/rspa.1998.0167}{https://doi.org/10.1098/rspa.1998.0167} (1998).

\bibitem{threshold3} 
D. S. Wang, A. G. Fowler, A. M. Stephens, and L. C. L. Hollenberg
\newblock Threshold error rates for the toric and planar codes.
\newblock Quantum Info. Comput. 10, 456–469, \href{http://doi.org/10.26421/QIC10.5-6-6}{http://doi.org/10.26421/QIC10.5-6-6} (2010).

\bibitem{MWPM} 
J. Kelly, R. Barends, A. G. Fowler, A. Megrant, E. Jeffrey, T. C. White, D. Sank, J. Y. Mutus, B. Campbell, Y. Chen, et al.
\newblock State preservation by repetitive error detection in a superconducting quantum circuit.
\newblock Nature 519, 66, \href{https://doi.org/10.1038/nature14270}{https://doi.org/10.1038/nature14270} (2015).

\bibitem{google_qec1}
Google Quantum AI.
\newblock Exponential suppression of bit or phase errors with cyclic error correction.
\newblock Nature 595, 383, \href{https://doi.org/10.1038/s41586-022-05434-1}{https://doi.org/10.1038/s41586-022-05434-1} (2021).

\bibitem{repetition_flag}
Kim, Younghun, et al. 
\newblock Effectiveness of the syndrome extraction circuit with flag qubits on IBM quantum hardware.
\newblock Preprint at \href{https://arxiv.org/abs/2403.10217}{https://arxiv.org/abs/2403.10217} (2024).

\bibitem{ibm_repetition}
Wootton, James R. 
\newblock Benchmarking near-term devices with quantum error correction.
\newblock Quantum Science and Technology 5.4, 044004, \href{https://doi.org/10.1088/2058-9565/aba038}{https://doi.org/10.1088/2058-9565/aba038} (2020).

\bibitem{google_qec2} 
Google Quantum AI.
\newblock Suppressing quantum errors by scaling a surface code logical qubit. 
\newblock Nature 614, 676–681, \href{https://doi.org/10.1038/s41586-022-05434-1}{https://doi.org/10.1038/s41586-022-05434-1} (2023).

\bibitem{d2_qec} 
Andersen, C.K., Remm, A., Lazar, S. et al. 
\newblock Repeated quantum error detection in a surface code.
\newblock Nature Physics 16.8, 875-880, \href{https://doi.org/10.1038/s41567-020-0920-y}{https://doi.org/10.1038/s41567-020-0920-y} (2020).

\bibitem{d3_surf} 
Krinner, S., Lacroix, N., Remm, A. et al.
\newblock Realizing repeated quantum error correction in a distance-three surface code. 
\newblock Nature 605.7911, 669-674, \href{https://doi.org/10.1038/s41586-022-04566-8}{https://doi.org/10.1038/s41586-022-04566-8} (2022).

\bibitem{d3_surper} 
Zhao, Youwei, et al. 
\newblock Realization of an error-correcting surface code with superconducting qubits.
\newblock Phys. Rev. Lett. 129.3, 030501, \href{https://doi.org/10.1103/PhysRevLett.129.030501}{https://doi.org/10.1103/PhysRevLett.129.030501} (2022).

\bibitem{hardware_err}
Harper, Robin, and Steven T. Flammia. 
\newblock Learning correlated noise in a 39-qubit quantum processor. 
\newblock PRX Quantum 4.4, 040311, \href{https://doi.org/10.1103/PRXQuantum.4.040311}{https://doi.org/10.1103/PRXQuantum.4.040311} (2023).

\bibitem{d3_ion} 
Ryan-Anderson, Ciaran, et al. 
\newblock Realization of real-time fault-tolerant quantum error correction. 
\newblock Phys. Rev. X. 11.4, 041058, \href{https://doi.org/10.1103/PhysRevX.11.041058}{https://doi.org/10.1103/PhysRevX.11.041058} (2021).

\bibitem{d3_ion2} 
Nigg, Daniel, et al. 
\newblock Quantum computations on a topologically encoded qubit.
\newblock Science 345.619, 302-305, \href{https://doi.org/10.1126/science.1253742}{https://doi.org/10.1126/science.1253742} (2014).

\bibitem{outlook}
Devoret, Michel H., and Robert J. Schoelkopf.
\newblock Superconducting circuits for quantum information: an outlook.
\newblock Science 339.6124, 1169-1174, \href{https://doi.org/10.1126/science.1231930}{https://doi.org/10.1126/science.1231930} (2013).

\bibitem{universality}
Barenco, Adriano, et al. 
\newblock Elementary gates for quantum computation.
\newblock Physical review A 52.5, 3457, \href{https://doi.org/10.1103/PhysRevA.52.3457}{https://doi.org/10.1103/PhysRevA.52.3457} (1995).

\bibitem{universality2}
Boykin, P. Oscar, et al. 
\newblock A new universal and fault-tolerant quantum basis.
\newblock Information Processing Letters 75.3, 101-107, \href{https://doi.org/10.1016/S0020-0190(00)00084-3}{https://doi.org/10.1016/S0020-0190(00)00084-3} (2000).

\bibitem{universal}
Campbell, E. T., Terhal, B. M., and Vuillot, C. 
\newblock Roads towards fault-tolerant universal quantum computation. 
\newblock Nature 549(7671), 172-179, \href{https://doi.org/10.1038/nature23460}{https://doi.org/10.1038/nature23460} (2017).

\bibitem{transversal_restrict}
Eastin, B., and Knill, E.
\newblock Restrictions on transversal encoded quantum gate sets. 
\newblock Phys. Rev. Lett., 102.11, 110502, \href{https://doi.org/10.1103/PhysRevLett.102.110502}{https://doi.org/10.1103/PhysRevLett.102.110502} (2009).

\bibitem{transversality}
Zeng, Bei, Andrew Cross, and Isaac L. Chuang. 
\newblock Transversality versus universality for additive quantum codes.
\newblock IEEE Transactions on Information Theory 57.9, 6272-6284, \href{https://doi.org/10.1109/TIT.2011.2161917}{https://doi.org/10.1109/TIT.2011.2161917} (2011).

\bibitem{transversality2}
Kubica, Aleksander, and Michael E. Beverland. 
\newblock Universal transversal gates with color codes: A simplified approach. 
\newblock Physical Review A 91.3, 032330, \href{https://doi.org/10.1103/PhysRevA.91.032330}{https://doi.org/10.1103/PhysRevA.91.032330} (2015).

\bibitem{transversality3}
Jochym-O’Connor, Tomas, and Raymond Laflamme. 
\newblock Using concatenated quantum codes for universal fault-tolerant quantum gates. 
\newblock Physical review letters 112.1, 010505, \href{https://doi.org/10.1103/PhysRevLett.112.010505}{https://doi.org/10.1103/PhysRevLett.112.010505} (2014).

\bibitem{transversality4}
Anderson, Jonas T., Guillaume Duclos-Cianci, and David Poulin. 
\newblock Fault-tolerant conversion between the steane and reed-muller quantum codes.
\newblock Physical review letters 113.8, 080501, \href{https://doi.org/10.1103/PhysRevLett.113.080501}{https://doi.org/10.1103/PhysRevLett.113.080501} (2014).

\bibitem{trapped_ion_transversal}
Postler, L., Heu$\beta$en, S., Pogorelov, I. et al.
\newblock Demonstration of fault-tolerant universal quantum gate operations.
\newblock Nature 605.7911, 675-680, \href{https://doi.org/10.1038/s41586-022-04721-1}{https://doi.org/10.1038/s41586-022-04721-1}(2022).

\bibitem{entangling_gate}
Ryan-Anderson, C., et al. 
\newblock Implementing fault-tolerant entangling gates on the five-qubit code and the color code.
\newblock Preprint at \href{https://arxiv.org/abs/2208.01863}{https://arxiv.org/abs/2208.01863} (2022).

\bibitem{ibm_transversal}
Menendez, Daniel Honciuc, Annie Ray, and Michael Vasmer. 
\newblock Implementing fault-tolerant non-Clifford gates using the [[8,3,2]] color code. 
\newblock Preprint at \href{https://arxiv.org/abs/2309.08663}{https://arxiv.org/abs/2309.08663} (2023).

\bibitem{ion_cnot}
Bluvstein, D., Evered, S.J., Geim, A.A. et al.
\newblock Logical quantum processor based on reconfigurable atom arrays.
\newblock Nature 626.7997, 58-65, \href{https://doi.org/10.1038/s41586-023-06927-3}{https://doi.org/10.1038/s41586-023-06927-3} (2024).

\bibitem{carbon}
Da Silva, M. P., et al. 
\newblock Demonstration of logical qubits and repeated error correction with better-than-physical error rates.
\newblock Preprint at \href{https://arxiv.org/abs/2404.02280}{https://arxiv.org/abs/2404.02280} (2024).

\bibitem{heavy-hex-entangle}
B. Hetényi, J. R. Wootton. 
\newblock Creating entangled logical qubits in the heavy-hex lattice with topological codes.
\newblock Preprint at \href{https://arxiv.org/abs/2404.15989}{https://arxiv.org/abs/2404.15989} (2024).

\bibitem{teleportation}
Ryan-Anderson, C., et al. 
\newblock High-fidelity and Fault-tolerant Teleportation of a Logical Qubit using Transversal Gates and Lattice Surgery on a Trapped-ion Quantum Computer.
\newblock Preprint at \href{https://arxiv.org/abs/2404.16728}{https://arxiv.org/abs/2404.16728} (2024).

\bibitem{qme}
Gehér, György P., et al. 
\newblock Error-corrected Hadamard gate simulated at the circuit level.
\newblock Preprint at \href{https://arxiv.org/abs/2312.11605}{https://arxiv.org/abs/2312.11605} (2023).

\bibitem{torino}
McKay, David C., et al. 
\newblock Benchmarking quantum processor performance at scale.
\newblock Preprint at \href{https://arxiv.org/abs/2311.05933}{https://arxiv.org/abs/2311.05933} (2023).

\bibitem{syndrom_extraction}
D. P. DiVincenzo and P. W. Shor.
\newblock Fault-tolerant error correction with efficient quantum codes.
\newblock Phys. Rev. Lett. 77, 3260, \href{https://doi.org/10.1103/PhysRevLett.77.3260}{https://doi.org/10.1103/PhysRevLett.77.3260} (1996).

\bibitem{syndrom_extraction_2}
A. M. Steane.
\newblock Active stabilization, quantum computation, and quantum state synthesis.
\newblock Phys. Rev. Lett. 78, 2252, \href{https://doi.org/10.1103/PhysRevLett.78.2252}{https://doi.org/10.1103/PhysRevLett.78.2252} (1997).

\bibitem{syndrom_extraction_3}
E. Knill.
\newblock Quantum computing with realistically noisy devices.
\newblock Nature 434, 39, \href{https://doi.org/10.1038/nature03350}{https://doi.org/10.1038/nature03350} (2005).

\bibitem{qec_guide} 
Roffe, Joschka. 
\newblock Quantum error correction: an introductory guide.
\newblock Contemporary Physics 60.3, 226-245, \href{https://doi.org/10.1080/00107514.2019.1667078}{https://doi.org/10.1080/00107514.2019.1667078} (2019).

\bibitem{flag1}
R. Chao and B. W. Reichardt.
\newblock Quantum error correction with only two extra qubits.
\newblock Phys. Rev. Lett. 121, 050502, \href{https://doi.org/10.1103/PhysRevLett.121.050502}{https://doi.org/10.1103/PhysRevLett.121.050502} (2018).

\bibitem{flag2}
R. Chao and B. W. Reichardt.
\newblock Flag fault-tolerant error correction for any stabilizer code.
\newblock PRX Quantum 1, 010302, \href{https://doi.org/10.1103/PRXQuantum.1.010302}{https://doi.org/10.1103/PRXQuantum.1.010302} (2020).

\bibitem{ibm_qec1}
Chen, E. H. et al. 
\newblock Calibrated decoders for experimental quantum error correction. 
\newblock Phys. Rev. Lett. 128, 110504, \href{https://doi.org/10.1103/PhysRevLett.128.110504}{https://doi.org/10.1103/PhysRevLett.128.110504} (2022).

\bibitem{ibm_qec2}
Sundaresan, N., Yoder, T.J., Kim, Y. et al. 
\newblock Demonstrating multi-round subsystem quantum error correction using matching and maximum likelihood decoders. 
\newblock Nat. Commun. 14, 2852, \href{https://doi.org/10.1038/s41467-023-38247-5}{https://doi.org/10.1038/s41467-023-38247-5} (2023).

\bibitem{hh-code}
Kim, Y., Kang, J., Kwon, Y. 
\newblock Design of quantum error correcting code for biased error on heavy-hexagon structure.
\newblock Quantum Information Processing, 22(6), 230, \href{https://doi.org/10.1007/s11128-023-03979-2}{https://doi.org/10.1007/s11128-023-03979-2} (2023).

\bibitem{heavyhexagon}
C. Chamberland, G. Zhu, T. J. Yoder, J. B. Hertzberg, and A. W. Cross
\newblock Topological and subsystem codes on low-degree graphs with flag qubits.
\newblock Phys. Rev. X 10, 011022, \href{https://doi.org/10.1103/PhysRevX.10.011022}{https://doi.org/10.1103/PhysRevX.10.011022} (2020).

\bibitem{strategy}
Benito, César, et al. 
\newblock Comparative study of quantum error correction strategies for the heavy-hexagonal lattice.
\newblock Preprint at \href{https://arxiv.org/abs/2402.02185}{https://arxiv.org/abs/2402.02185} (2024).

\bibitem{correlated_err}
Nickerson, Naomi H., and Benjamin J. Brown. 
\newblock Analysing correlated noise on the surface code using adaptive decoding algorithms. 
\newblock Quantum 3, 131, \href{https://doi.org/10.22331/q-2019-04-08-131}{https://doi.org/10.22331/q-2019-04-08-131} (2019).

\bibitem{leakage_err}
McEwen, Matt, et al. 
\newblock Removing leakage-induced correlated errors in superconducting quantum error correction.
\newblock Nature communications 12.1, 1761, \href{https://doi.org/10.1038/s41467-021-21982-y}{https://doi.org/10.1038/s41467-021-21982-y} (2021).

\bibitem{decoding_cnot}
Cain, Madelyn, et al.
\newblock Correlated decoding of logical algorithms with transversal gates.
\newblock Preprint at \href{https://arxiv.org/abs/2403.03272}{https://arxiv.org/abs/2403.03272} (2024).

\bibitem{correlated}
Paler, A., and Fowler, A. G.
\newblock Pipelined correlated minimum weight perfect matching of the surface code. 
\newblock Quantum 7, 1205, \href{https://doi.org/10.22331/q-2023-12-12-1205}{https://doi.org/10.22331/q-2023-12-12-1205} (2023).

\bibitem{topology_MWPM} 
Eric Dennis, Alexei Kitaev, Andrew Landahl, and John Preskill.
\newblock Topological quantum memory.
\newblock Journal of Mathematical Physics, 43(9):4452–4505, \href{https://doi.org/10.1063/1.1499754}{https://doi.org/10.1063/1.1499754} (2002).

\bibitem{subsystem_MWPM}
Sergey Bravyi, Guillaume Duclos-Cianci, David Poulin, and Martin Suchara. 
\newblock Subsystem surface codes with three-qubit check operators. 
\newblock Preprint at \href{https://arxiv.org/abs/1207.1443}{https://arxiv.org/abs/1207.1443} (2012).

\bibitem{XZZX_MWPM}
Bonilla Ataides, J. Pablo, et al.  
\newblock The XZZX surface code. 
\newblock Nature communications 12.1, 2172, \href{https://arxiv.org/abs/1207.1443}{https://arxiv.org/abs/1207.1443} (2021).

\bibitem{MWPM_2}
A. G. Fowler, A. C. Whiteside, and L. C. Hollenberg.
\newblock Towards practical classical processing for the surface code.
\newblock Physical review letters 108, 180501, \href{https://doi.org/10.1103/PhysRevLett.108.180501}{https://doi.org/10.1103/PhysRevLett.108.180501} (2012).

\bibitem{MWPM_3}
Fowler, Austin G.
\newblock Minimum weight perfect matching of fault-tolerant topological quantum error correction in average $ O (1) $ parallel time. 
\newblock Quantum Information \& Computation 15, 145, \href{http://doi.org/10.26421/QIC15.1-2-9}{http://doi.org/10.26421/QIC15.1-2-9} (2015).

\bibitem{stim}
C, Gidney
\newblock Stim: a fast stabilizer circuit simulator.
\newblock Quantum 5, 497, \href{https://doi.org/10.22331/q-2021-07-06-497}{https://doi.org/10.22331/q-2021-07-06-497} (2021).

\bibitem{pymatching}
O, Higgott
\newblock PyMatching: A Python Package for Decoding Quantum Codes with Minimum-Weight Perfect Matching.
\newblock ACM Transactions on Quantum Computing 3.3, \href{https://doi.org/10.1145/3505637}{https://doi.org/10.1145/3505637} (2022)

\bibitem{pymatching2}
O, Higgott, C, Gidney
\newblock Sparse Blossom: correcting a million errors per core second with minimum-weight matching.
\newblock Preprint at \href{https://arxiv.org/abs/2303.15933}{https://arxiv.org/abs/2303.15933} (2023)

\bibitem{qiskit}
IBM Quantum and Community.
\newblock Qiskit: An open-source framework for quantum computing.
\newblock \href{https://doi.org/10.5281/zenodo.2573505}{https://doi.org/10.5281/zenodo.2573505} (2021).

\bibitem{ANN_Dec}
Spiro Gicev, Lloyd Hollenberg, Muhammad Usman.
\newblock A fast and scalable artificial neural network syndrome decoder for surface codes.
\newblock Quantum 7, 1058, \href{https://doi.org/10.22331/q-2023-07-12-1058}{https://doi.org/10.22331/q-2023-07-12-1058} (2023).

\bibitem{ANN_IBM}
Brhyeton Hall, Spiro Gicev, Muhammad Usman.
\newblock Artificial neural network syndrome decoding on IBM Quantum Processors.
\newblock Preprint at \href{https://arxiv.org/abs/2311.15146}{https://arxiv.org/abs/2311.15146} (2023).

\bibitem{google_RNN}
Bausch, Johannes, et al. 
\newblock Learning to decode the surface code with a recurrent, transformer-based neural network.
\newblock Preprint at \href{https://arxiv.org/abs/2310.05900}{https://arxiv.org/abs/2310.05900} (2023).

\bibitem{ECR}
Sundaresan, Neereja, et al. 
\newblock Reducing unitary and spectator errors in cross resonance with optimized rotary echoes. 
\newblock PRX Quantum 1.2, 020318, \href{https://doi.org/10.1103/PRXQuantum.1.020318}{https://doi.org/10.1103/PRXQuantum.1.020318} (2020).

\bibitem{DD}
Tripathi, Vinay, et al. 
\newblock Suppression of crosstalk in superconducting qubits using dynamical decoupling.
\newblock Physical Review Applied 18.2, 024068, \href{https://doi.org/10.1103/PhysRevApplied.18.024068}{https://doi.org/10.1103/PhysRevApplied.18.024068} (2022).

\end{thebibliography}
\end{document}